\documentclass[%
reprint,
twocolumn,
superscriptaddress,
nofootinbib,
 amsmath,amssymb,
 aps,
]{revtex4-1}
\usepackage{graphicx}
\usepackage{dcolumn}
\usepackage{array}
\usepackage{hyperref}
\usepackage{bm}
\usepackage{algorithm2e}
\usepackage{xcolor}
\usepackage{enumitem}
\usepackage{graphicx}
\usepackage[export]{adjustbox}
\usepackage{units}

\graphicspath{{figures/}}
\begin{document}

\newcommand{\xxf}[1]{\textcolor{purple}{{\it{[XXK: #1]}}}}
\newcommand{\sa}[1]{\textcolor{blue}{{\it{[SA: #1]}}}}
\newcommand{\changed}[1]{\textcolor{black}{{{#1}}}}


\preprint{APS/123-QED}

\title{Progress toward the detection of the gravitational-wave background from stellar-mass binary black holes: a mock data challenge}

\author{Xiao-Xiao Kou}
\email{kou00016@umn.edu}
\affiliation{%
University of Minnesota, School of Physics and Astronomy, Minneapolis, MN 55455, USA}%
\author{Muhammed Saleem}
\email{muhammed.cholayil@austin.utexas.edu}
\affiliation{%
University of Minnesota, School of Physics and Astronomy, Minneapolis, MN 55455, USA}%
\affiliation{Center for Gravitational Physics, University of Texas at Austin, Austin, TX 78712, USA}
\author{Vuk Mandic}
\affiliation{%
University of Minnesota, School of Physics and Astronomy, Minneapolis, MN 55455, USA}%
\author{Colm Talbot}
\affiliation{Kavli Institute for Cosmological Physics, University of Chicago, USA}
\author{Eric Thrane}
\affiliation{School of Physics and Astronomy, Monash University, VIC 3800, Australia}
\affiliation{OzGrav: The ARC Centre of Excellence for Gravitational-Wave Discovery, Clayton, VIC 3800, Australia}


\begin{abstract}
While the third LIGO--Virgo gravitational-wave transient catalog includes 90 signals, it is believed that ${\cal O}(10^5)$ binary black holes merge somewhere in the Universe every year.
Although these signals are too weak to be detected individually with current observatories, they combine to create a stochastic background, which is potentially detectable in the near future.
LIGO--Virgo searches for the gravitational-wave background using cross-correlation have so far yielded upper limits.
However, Smith \& Thrane (2017) showed that a vastly more sensitive ``coherent'' search can be carried out by incorporating information about the phase evolution of binary black hole signals.
This improved sensitivity comes at a cost; the coherent method is computationally expensive and requires a far more detailed understanding of systematic errors than is required for the cross-correlation search. 
In this work, we demonstrate the coherent approach with realistic data, paving the way for a gravitational-wave background search with unprecedented sensitivity.
\end{abstract}

\maketitle

\section{Introduction}
The first three observing runs of the Advanced LIGO~\cite{LIGOScientific:2014pky} and Advanced Virgo~\cite{VIRGO:2014yos} gravitational wave (GW) detectors have revealed about 90 compact binary merger events ~\cite{LIGOScientific:2020ibl,LIGOScientific:2021djp,LIGOScientific:2021usb,KAGRA:2021kbb}. 
These observations have significantly contributed to our
understanding of stellar evolution and the formation of binary black holes (BBH)~\cite{Kimball:2020qyd,Zevin:2020gbd,Mapelli:2019bnp,Baibhav:2019gxm,Broekgaarden:2021efa,vanSon:2021zpk}.
In excess of two hundred additional binary black hole triggers have been observed during the first two phases of the fourth observing runs (O4a and O4b). 
However, these events are only a small fraction of the ${\cal O}(10^5)$ binary black hole mergers believed to take place in the Universe every year~\cite{KAGRA:2021duu}.
These unresolved mergers include contributions from high-redshift binaries, resulting in a gravitational-wave background that carries complementary information to existing, individually resolved observations of low-redshift binaries~\cite{Callister:2020arv,Smith:2020lkj,Turbang:2023tjk}. 

Gravitational-wave backgrounds can be characterized by their gravitational-wave energy density spectra:
\begin{equation}
\label{eq:omega}
\Omega_{\mathrm{gw}}(f) \equiv \frac{1}{\rho_c} \frac{\mathrm{d} \rho_{\mathrm{gw}}}{\mathrm{d} \ln f},
\end{equation}
where $\mathrm{d} \rho_{\mathrm{gw}}$ is the present-day gravitational-wave energy density between frequency $f$ and $f + \mathrm{d}f$ and $\rho_{c}$ is the critical energy density required to close the Universe. 
For stellar-mass binaries, in the frequency band accessible to terrestrial GW detectors (10-200 Hz) we expect the energy density spectrum to follow the spectrum predicted for a gravitational-wave driven inspiral:
\begin{align}
    \Omega_\text{gw}(f) \propto f^{2/3} .
\end{align}

Some potential sources of the gravitational-wave background are expected to be Gaussian and are therefore fully described in terms of their energy density spectra. The optimal search strategy for such models is the cross-correlation approach~\cite{Allen:1997ad} that quantifies the level of correlation in multiple detectors. 
Data collected during the first three observing runs of Advanced LIGO and Advanced Virgo has been utilized to search for the GW background with the cross-correlation technique. So far, only upper limits have been established~\cite{KAGRA:2021kbb,LIGOScientific:2019vic,LIGOScientific:2018mvr,LIGOScientific:2016jlg}, with the most stringent upper limit from Ref.~\cite{KAGRA:2021kbb} setting $\Omega_{\mathrm{GW}} \leq 3.4\times 10^{-9}$ at $\unit[25]{Hz}$ for a power-law background with a spectral index of 2/3. 

However, the GW background due to BBH mergers is highly \textit{non-Gaussian} because the individual signals that contribute to it are sparse and rarely overlap in time and frequency. For a background to be approximately Gaussian, there would need to be a large number of overlapping events at every frequency and time—effectively ensuring that the signal is always present and varies smoothly. In contrast, given merger rate density estimates of $\mathcal{R}(z=0.2)=\unit[19-42]{Gpc^{-3},yr^{-1}}$\cite{KAGRA:2021duu}, BBH signals are present in just $\approx 1\%$ of the data above $\unit[15]{Hz}$. As a result, simultaneous overlaps of two or more signals are rare in the frequency band of terrestrial GW detectors~\cite{LIGOScientific:2017zlf,Johnson:2024foj}.

Smith \& Thrane~\cite{Smith:2017vfk} proposed a statistically optimal approach to search for and estimate the amplitude of the BBH background, making use of BBH waveform models to look for sub-threshold BBH signals.
In this approach, the data are divided into segments and each segments is assigned a probability of containing an astrophysical signal using Bayesian inference.
The segments are then combined to determine if a {\it population of sub-threshold BBH events} are present in the data.
By incorporating information about the BBH waveform, this ``phase coherent approach'' is expected to produce a detection $\gtrsim 1000\times$ faster than the cross-correlation analysis, which includes no information about the morphology of binary black hole signals.\footnote{Various other authors have explored methods of searching for non-Gaussian backgrounds including Refs.~\cite{Drasco,Martellini:2014xia,popcorn,Sharma:2020btq,Buscicchio:2022raf,Braglia:2022icu,Yamamoto:2022kuh,Lawrence:2023buo,Suvodeep:2019,Suvodeep:2023}.}

Since Ref.~\cite{Smith:2017vfk}, progress has been made on modeling and mitigating several important sources of systematic error, e.g.,~\cite{student-t,windows,Biscoveanu}. 
In this work we combine and extend these methods and report a robust implementation of the phase-coherent method proposed in Ref.~\cite{Smith:2017vfk}. 
We show that implementation of the phase-coherent approach is challenging, though possible, in practice.
The improved sensitivity of the search requires precise control of systematic errors. 
We show how to manage these systematic errors and demonstrate a fully functional pipeline on mock data, thereby paving the way for a phase-coherent search with real data.

The remainder of this paper is organized as follows. Section~\ref{sec:method} outlines the formalism underlying our analysis. Section~\ref{sec:systematics} discusses key sources of systematic errors, including uncertainties in the noise power spectral density and transient noise artifacts (glitches). In Section~\ref{sec:solution}, we present our approach to mitigating these systematics and demonstrate the full analysis framework with simulations. Section~\ref{sec:real-o3-noise} applies this framework to time-reversed LIGO O3b data with injected signals. Finally, Section~\ref{sec:summary} summarizes our findings and outlines directions for future work.

\section{Method}\label{sec:method}
\subsection{A single data segment}
\label{subsec:PE}

In this section, we describe the phase-coherent method for detecting a binary black hole background as introduced in Ref.~\cite{Smith:2017vfk}. For now, we deliberately set aside two important subtleties: uncertainty in the noise model and the finite-duration effects that arise when a stochastic process is truncated. While these issues were not addressed in the original work~\cite{Smith:2017vfk}, they were independently studied in later papers~\cite{student-t,windows}. Both effects introduce systematic errors, which we analyze in the subsequent sections.

The detector output is the strain time series that can be written as
\begin{align}
    d(t) = n(t) + h(t),
\end{align}
where $n(t)$ is the detector noise and $h(t)$ is the detector response to the gravitational-wave signal. The noise $n(t)$ is approximately Gaussian and stationary, though we consider non-Gaussian effects in a later section. 
Gaussian noise in a segment of duration $T$ is characterized by one-sided power spectral density (PSD)
\begin{align}
P(f) \equiv \frac{2}{T}\left\langle | \tilde{n}(f)|^2\right\rangle,
\end{align}
where the tilde denotes a discrete Fourier transform, $f$ is the frequency, and the angled brackets denote an ensemble average.

In the case of Gaussian noise, the data are distributed according to the Whittle likelihood~\cite{intro}:
\begin{equation}
\log \left[\mathcal{L}\left(d\,|\, \theta\right)\right] = -\frac{1}{2}\left\langle \tilde{d} - \tilde{h} \left(\theta\right), \tilde{d} - \tilde{h} \left(\theta\right) \right \rangle + \text{const} .
\label{eq:whittle}
\end{equation}
Here, the gravitational-wave signal $h$ depends on 15 binary black hole parameters denoted by $\theta$.
For compact notation, we employ the noise-weighted inner product
\begin{equation}
\langle a, b\rangle \equiv 4 \Re \Delta f \sum_k \frac{a^*\left(f_k\right) b\left(f_k\right)}{P\left(f_k\right)},
\label{eq:inner_product}
\end{equation}
where $\Delta f$ is the frequency-bin size and the sum runs over $k$ frequency bins.

\begin{figure*}[ht]
    \includegraphics[width=\textwidth,height=0.306\textwidth]{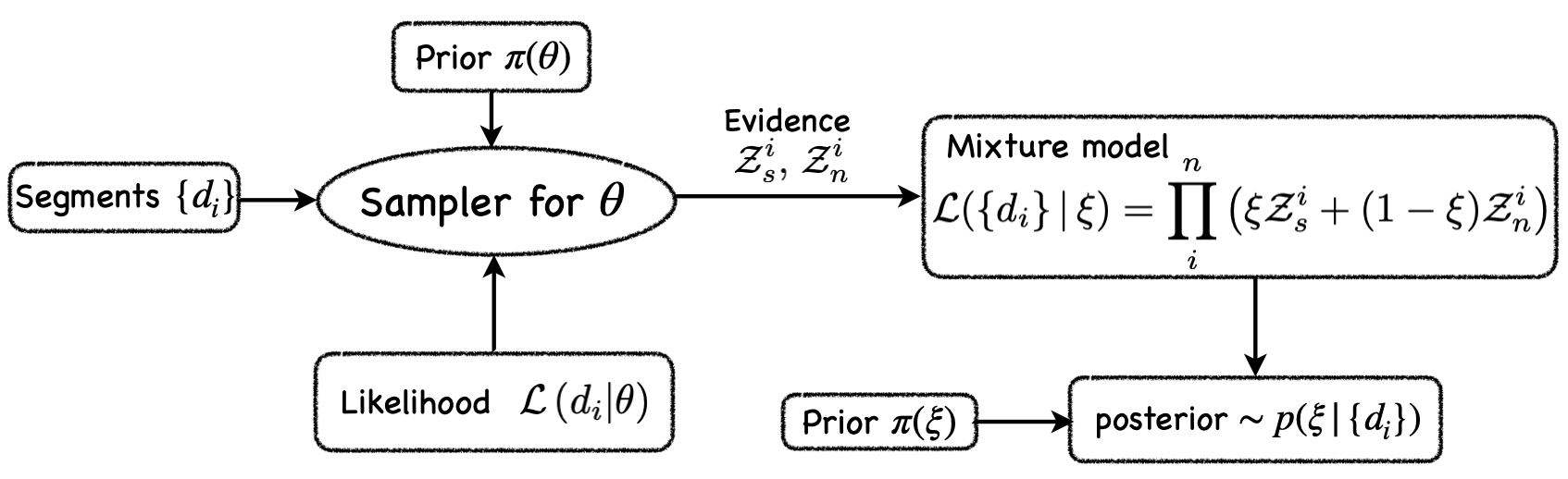}
    \caption{A flowchart depicting the analysis from Smith \& Thrane \cite{Smith:2017vfk}, which is reviewed here in Section.~\ref{sec:method}. 
    The data are divided into segments denoted $d_i$.
    Each segment is analyzed to determine the binary black hole parameters $\theta$.
    This produces two evidence values for each segment: ${\cal Z}_s$ for the signal hypothesis and ${\cal Z}_n$ for the noise hypothesis.
    The evidence values are input to a mixture model, which is used to estimate the duty cycle $\xi$, which is the fraction of segments containing a signal. Note that this flowchart does not include the main topics discussed in this work: the subtleties from noise uncertainties and the finite-duration effects.
    }
    \label{fig:general_tbs}
\end{figure*}

Following Bayes’ theorem, the posterior on $\theta$ given data $d$ is given by
\begin{equation}
\begin{aligned}
p(\theta\, | \, d) &=\frac{\mathcal{L}(d\, | \, \theta) \pi(\theta)}{\mathcal{Z}_s},
\label{eq:bayes}
\end{aligned}
\end{equation}
where $\pi(\theta)$ represents the prior probability distribution assumed for $\theta$ and $\mathcal{Z}_s$ is the  \textit{signal evidence}, also referred to as the marginal likelihood for the signal model, defined as 
\begin{equation}
\begin{aligned}
\mathcal{Z}_s(d) &= p(d) = \int \mathcal{L}(d\, | \, \theta) \pi(\theta) d \theta .
\label{eq:bayes-Zs}
\end{aligned}
\end{equation}
We also define the noise evidence, ${\cal Z}_n$, which is calculated assuming $h=0$,
\begin{equation}
\begin{aligned}
\mathcal{Z}_n(d) &= \mathcal{L}(d\, | \, h = 0).
\label{eq:bayes-Zn}
\end{aligned}
\end{equation}
Unlike the signal evidence, the noise evidence does not require the $\theta$-marginalization step of Eq.~\ref{eq:bayes-Zs}, and is straightforward to calculate using Eq.~\ref{eq:whittle}.
The posterior probability density function and the evidence are typically obtained using nested sampling~\cite{Skilling:2004pqw,Skilling:2006gxv}. 

The Bayesian evidence values ${\cal Z}_s$ and ${\cal Z}_n$ are the key ingredients for the phase-coherent search.
For a single segment, the presence of a binary black hole signal can be quantified by a Bayes factor comparing the signal evidence to the noise evidence:
\begin{align}
    \text{BF} = {\cal Z}_s / {\cal Z}_n .
\end{align}
In the next section we discuss how to combine observations from multiple segments.

\subsection{Combining multiple segments}
\label{sec:tbs_introduction}
The next step is to combine data from many segments, denoted $d_i$. 
The segment duration is chosen to be sufficiently long to include a complete high-mass binary black hole signal in the sensitive frequency band ($>\unit[20]{Hz}$ in this analysis) but short enough to make the occurrence of two such signals within a single segment unlikely. 
We employ a segment duration of $\unit[4]{s}$.

Following \cite{Smith:2017vfk}, we construct a mixture model likelihood for the segment $i$
\begin{equation}
\mathcal{L}\left(d_i\,|\, \theta, \xi\right)=\xi \mathcal{L}\left(d_i\,|\, \theta\right)+(1-\xi) \mathcal{L}\left(d_i\,|\, h=0\right) .
\label{eq:mixture}
\end{equation}
Here, $\xi$ is the probability that the segment contains a signal described by the signal likelihood ${\cal L}(d|\theta)$.
There is a probability $1-\xi$ that the segments contain noise described by the noise likelihood denoted $\mathcal{L}\left(d_i\,|\, h = 0\right)$.
We marginalize over the binary black hole parameter space $\theta$ to obtain:
\begin{equation}
\mathcal{L}\left(d_i\,|\, \xi\right)=\xi \mathcal{Z}_s^i + (1-\xi) \mathcal{Z}_n^i ,
\label{eq:simple_mixture}
\end{equation}
where we have incorporated the notations introduced in Eq.~\ref{eq:bayes-Zs} and \ref{eq:bayes-Zn}.

Combining $N$ independent data segments, the likelihood for the entire dataset $\mathcal{D}$ becomes
\begin{equation}
\mathcal{L}\left(\mathcal{D}\,|\, \xi \right) = \prod_i^N \left( \xi Z_s^i + (1-\xi) Z_n^i \right).
\label{eq:total_mixture}
\end{equation}
The variable $\xi$ can now be interpreted as a ``duty cycle'' parameter equal to the fraction of segments that include a BBH signal. Given the number of segments analyzed, it is straightforward to convert the duty cycle into binary black hole merger rates as discussed in~\cite{Smith:2017vfk}.
The posterior for the duty cycle is 
\begin{equation}
p(\xi\,|\,\mathcal{D}) \propto \mathcal{L} \left( \mathcal{D} \,|\, \xi \right) \pi(\xi),
\label{eq:total_xi}
\end{equation}
where $\pi(\xi)$ is the prior distribution on $\xi$, which we assume here to be uniform. 
If the posterior excludes $\xi=0$ with significant credibility, one may infer the presence of a binary black hole background in the data.
A flow chart illustrating the entire process is provided in Fig.~\ref{fig:general_tbs}.

The original work by Smith \& Thrane \cite{Smith:2017vfk} excluded resolvable events with matched-filter network signal-to-noise ratios $\text{SNR}\geq 12$ to ensure that the existence of a stochastic background was not simply inferred from the presence of resolvable binaries.
Ref.~\cite{Renzini} shows that relatively nearby binaries contribute a large fraction of the signal-to-noise ratio in a phase-coherent search.
However, Ref.~\cite{Smith:2020lkj} showed that phase coherent approach can distinguish between backgrounds that differ only at cosmological distances, well beyond the most distant events with network SNR=8; see their Fig.~4.
Thus, as data is collected and the signal-to-noise ratio grows, the phase coherent search provides information from increasingly distant binaries.\footnote{See also Ref.~\cite{Callister:2020arv} for a discussion of the complementarity of searches for resolved and unresolved binaries, albeit in the context of a cross-correlation search.}

Formally, we should modify the likelihood function to account for the ``reverse selection effect'' of removing resolvable binaries. 
The selection effect introduces a normalisation factor $p_\text{det}$ to account for the fact that some data is removed~\cite{intro,Essick:2023upv}:
\begin{align}\label{eq:pdet}
    \mathcal{L}(d_i | \xi) 
    &= \frac{\xi \mathcal{Z}_s^i + (1-\xi) \mathcal{Z}_n^i}{p_{\rm det}(\xi)} .
\end{align}
We describe how to calculate $p_\text{det}$ in Appendix~\ref{appendix:selection-effects}.
We show that the exclusion of resolvable binaries with network $\text{SNR}>12$ only slightly widens our posterior for $\xi$---consistent with results from Ref.~\cite{Smith:2017vfk}.

\section{Systematic errors}\label{sec:systematics}
\subsection{Sources of systematic error}
The analysis framework described in Section~\ref{sec:method} is built by assuming certain ideal noise conditions, while the real noise is far from these assumptions. In this section, we discuss three subtleties that arise when the phase-coherent analysis is carried out in practice: uncertainty in the noise power spectral density, so-called finite-duration effects, and glitches. The first two of these add complexity not discussed in the original proposal by Smith \& Thrane~\cite{Smith:2017vfk}. 

\textit{Noise uncertainty.}
    In Eq.~\ref{eq:whittle} and ~\ref{eq:inner_product}, we implicitly assume perfect knowledge about the noise power spectral density $P(f)$. However, in a realistic analysis, $P(f)$ is \textit{estimated} from finite amount of data, and this estimate is necessarily uncertain. Several studies have investigated how to incorporate this uncertainty into gravitational-wave inference calculations~\cite{Banagiri:2019lon,student-t,Biscoveanu}.  Failing to take into account noise uncertainty leads to bias in the analysis of single events, comparable to systematic error from waveform systematics and calibration uncertainty \cite{student-t,Biscoveanu}. However, when many events are combined together as we seek to do here, the accumulated error can significantly bias our estimate of duty cycle.

\textit{Finite duration effects.}
    Another implicit assumption in  Eq.~\ref{eq:total_xi} is that the detector noise is generated with periodic boundary conditions. This assumption is almost always used in gravitational-wave data analysis because it means that the noise covariance matrix in the Whittle likelihood is diagonal (as in Eq.~\ref{eq:whittle})~\cite{Allen:2001ay,1984Unser}. In practice, however, the noise does not have periodic boundary conditions. The window functions used to truncate the data into finite durations introduce off-diagonal elements to the noise covariance matrix \cite{windows,Isi:2021iql,Siegel:2024jqd,Burke:2025bun}. Failing to take into account these so-called finite-duration effects leads to systematic errors comparable to those from uncertainty in the noise estimate.
    Ref. \cite{windows} demonstrated that finite-duration effects can lead to bias in phase-coherent searches for gravitational-wave backgrounds.
    
\textit{Glitches.} 
    The third source of systematics is non-Gaussian transient noise, also known as glitches.  While noise uncertainty and finite-duration effects are important even for the analysis of idealized Gaussian noise, glitches are a feature of real GW detector data, and they can mimic the presence of true GW signals.
    Without mitigation, they introduce significant biases in the duty cycle posterior, as shown in~\cite{Smith:2017vfk}. Thus, modeling non-Gaussian noise is essential to prevent glitches from being misidentified as gravitational-wave signals. 
    We employ the glitch model from \cite{Smith:2017vfk}.

\begin{figure}[t]
\centering
\includegraphics[width=0.49\textwidth,left]{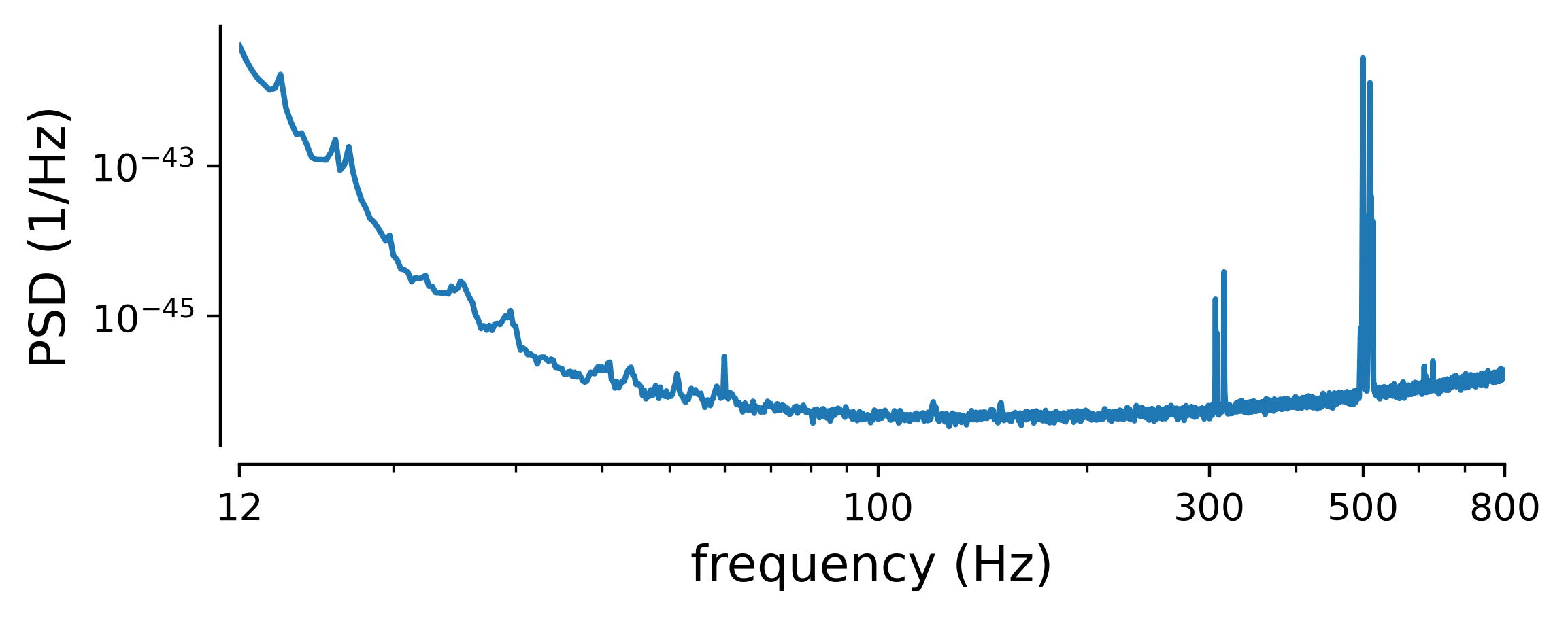}
\includegraphics[width=0.49\textwidth,left]{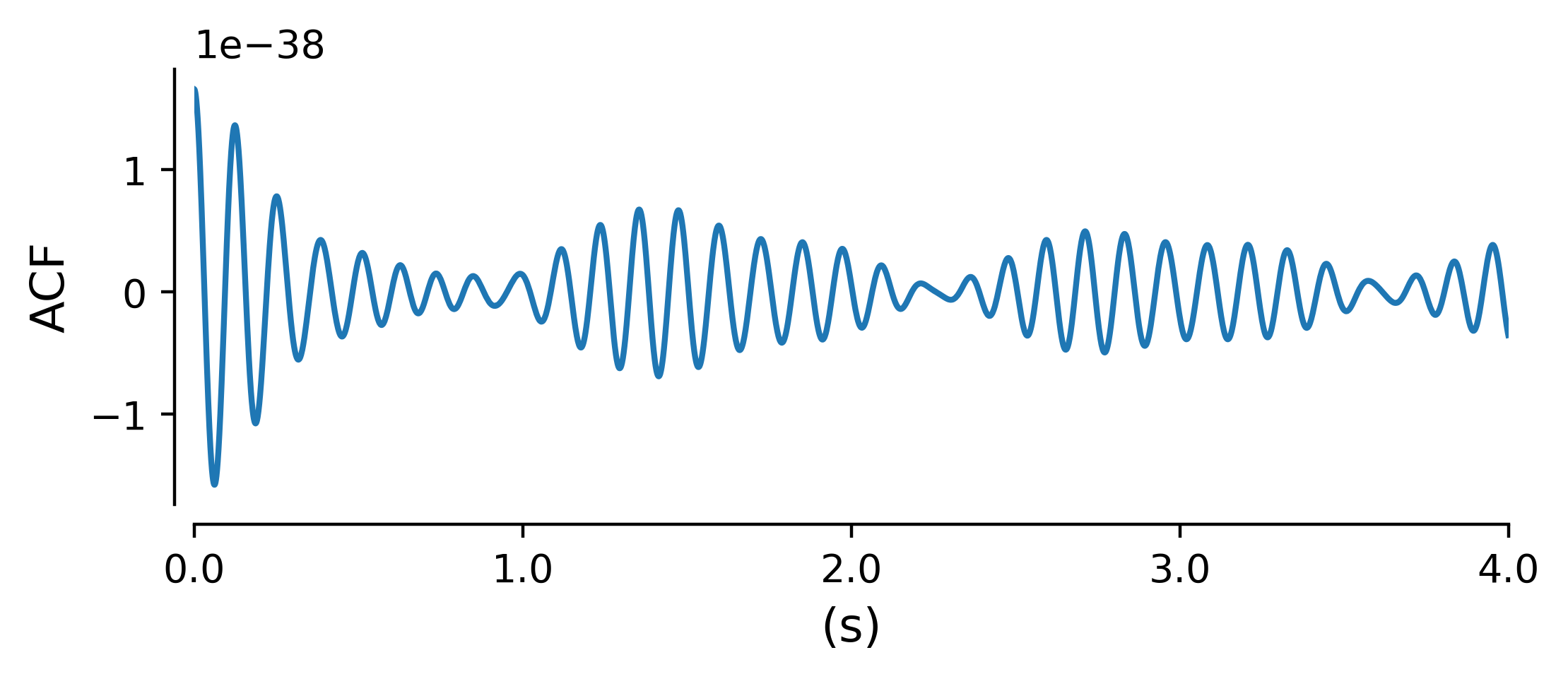}
\includegraphics[width=0.49\textwidth,left]{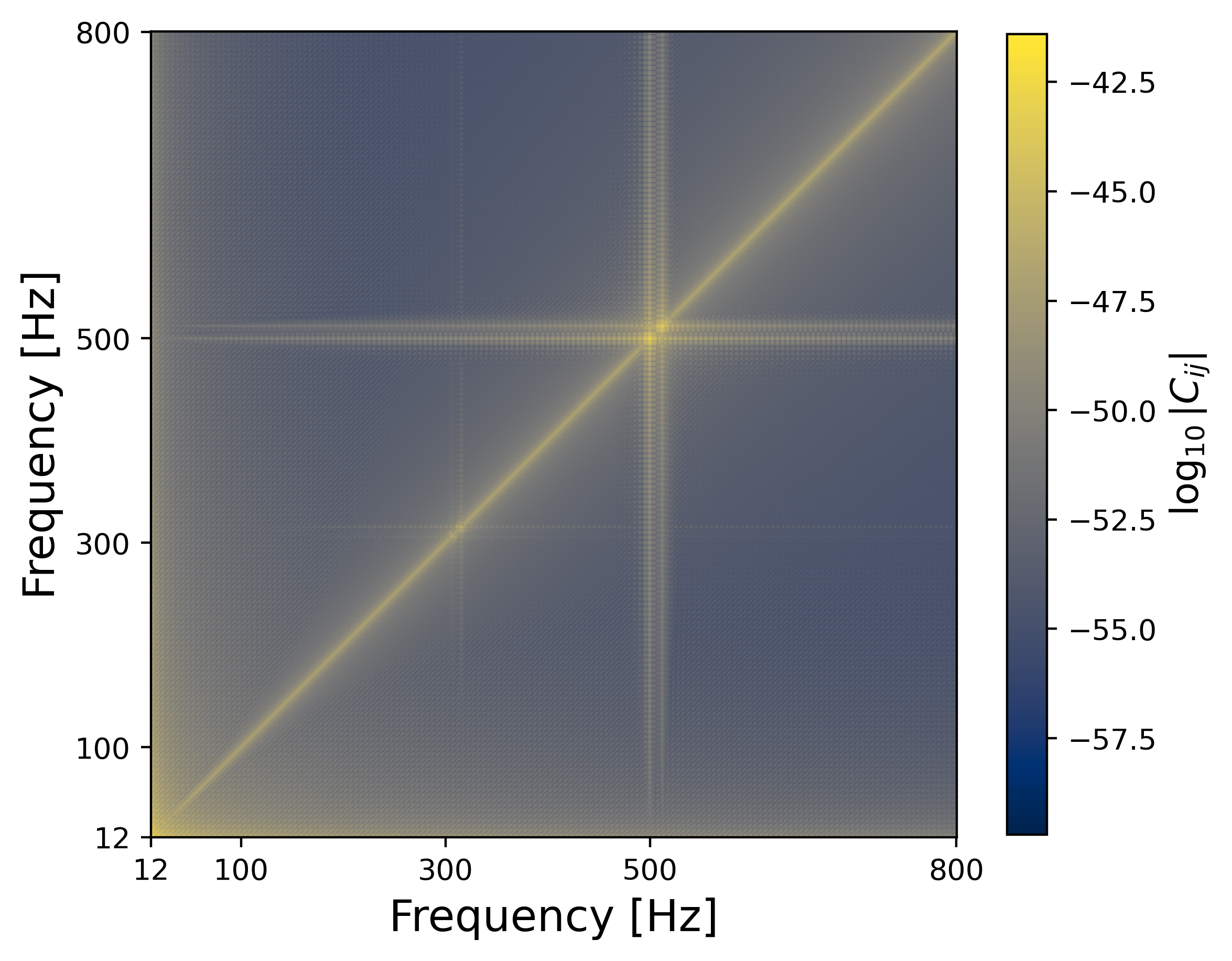}
\caption{
Characterizing LIGO noise.
Top: the noise power spectral density from LIGO Livingston data collected around the time of the binary black hole merger GW170814, estimated using techniques introduced by \cite{windows}. 
Middle: the implied auto-correlation function associated with this PSD. 
Bottom: Reproduction of the top-left panel of Fig. 12 from \cite{windows}, showing the associated covariance matrix computed using the Tukey window and following the algorithm detailed therein.}
\label{fig:psd_acf_cij}
\end{figure}

\subsection{Simulations: frequency-domain and time-domain datasets}\label{sec:simulations}
In order to demonstrate the importance of noise uncertainty and finite-duration effects, we simulate two different datasets: the ``FD dataset'' and the ``TD datasets''. \\

\textbf{Frequency domain (FD) datasets:}
The defining feature of an FD (frequency-domain) dataset is that it is directly simulated in the frequency domain using periodic boundary conditions. This approach follows the same method used by Smith \& Thrane~\cite{Smith:2017vfk} for their simulated dataset. In our case, we generate Gaussian noise in the frequency domain based on the PSD shown in Fig.~\ref{fig:psd_acf_cij} (top), which is derived from real data near the time of the binary black hole merger GW170814~\cite{LIGOScientific:2017ycc}. Since the data are not truncated, periodic boundary conditions are naturally enforced, resulting in frequency bins that are approximately statistically independent. As a consequence, the noise covariance matrix is approximately diagonal in the frequency domain.

We generate GW signals using the \texttt{IMRPhenomPv2} approximant~\cite{Hannam:2013oca}. The source parameters are randomly drawn; the component masses in the detector frame are uniformly distributed on the interval \((30,50)M_{\odot}\), with a mass ratio between \((0.5,1)\). Dimensionless spin magnitudes \((\chi_1, \chi_2)\) are uniformly distributed in \((0, 0.8)\), within the domain of validity for \texttt{IMRPhenomPv2}. The spin unit vectors follow an isotropic distribution. Luminosity distances are drawn from a uniform in-comoving-volume distribution in the range \((0.10\,\mathrm{Gpc}, 5\,\mathrm{Gpc})\).
Binary orientation and sky location angles are sampled isotropically. The coalescence time \(t_c\) follows a uniform distribution centered around \(t_s + 3\) with a width of 1\,s, where \(t_s\) is the starting time of the $\unit[4]{s}$ segment.
We construct three data sets corresponding to \(\xi = (0, 0.05, 0.1)\), each containing 3000 segments.\\

\textbf{Time domain (TD) datasets:}
In contrast to FD datasets, a TD dataset is generated as a time series, which must be segmented and windowed before being Fourier transformed for analysis. This segmentation breaks the assumption of periodic boundary conditions and introduces correlations between frequency bins—an effect known as the finite-duration effect~\cite{windows}. As a result, the frequency-domain representation of a TD dataset has a noise covariance matrix with off-diagonal elements.

Moreover, in our TD simulations, we assume that the true noise PSD is not directly available for analysis. Instead, it must be estimated by averaging periodograms from neighboring data segments, introducing additional uncertainty into the PSD. Thus, compared to the idealized FD datasets, TD datasets introduce two key complications: correlated noise across frequency bins and uncertainty in the PSD—both of which are unavoidable in real gravitational-wave data analysis.

To simulate these effects, we follow the procedure of Ref.~\cite{student-t}. We begin by generating noise using the same PSD shown in Fig.\ref{fig:psd_acf_cij}, but with a total duration of $\unit[128]{s}$---much longer than the $\unit[4]{s}$ analysis segment. 
We then extract $\unit[4]{s}$ of data from the middle by truncating the time series on both sides, thereby eliminating periodic boundary conditions, and Fourier transforming the resulting segment. We follow the same injection distributions as used for the FD datasets.
As in the first simulation, we construct three datasets corresponding to \(\xi = (0, 0.05, 0.1)\), each containing 3000 segments.

\subsection{Demonstration}
\label{sec:demonstration}

\begin{figure}
    \centering
    \includegraphics[width=1.0\linewidth]{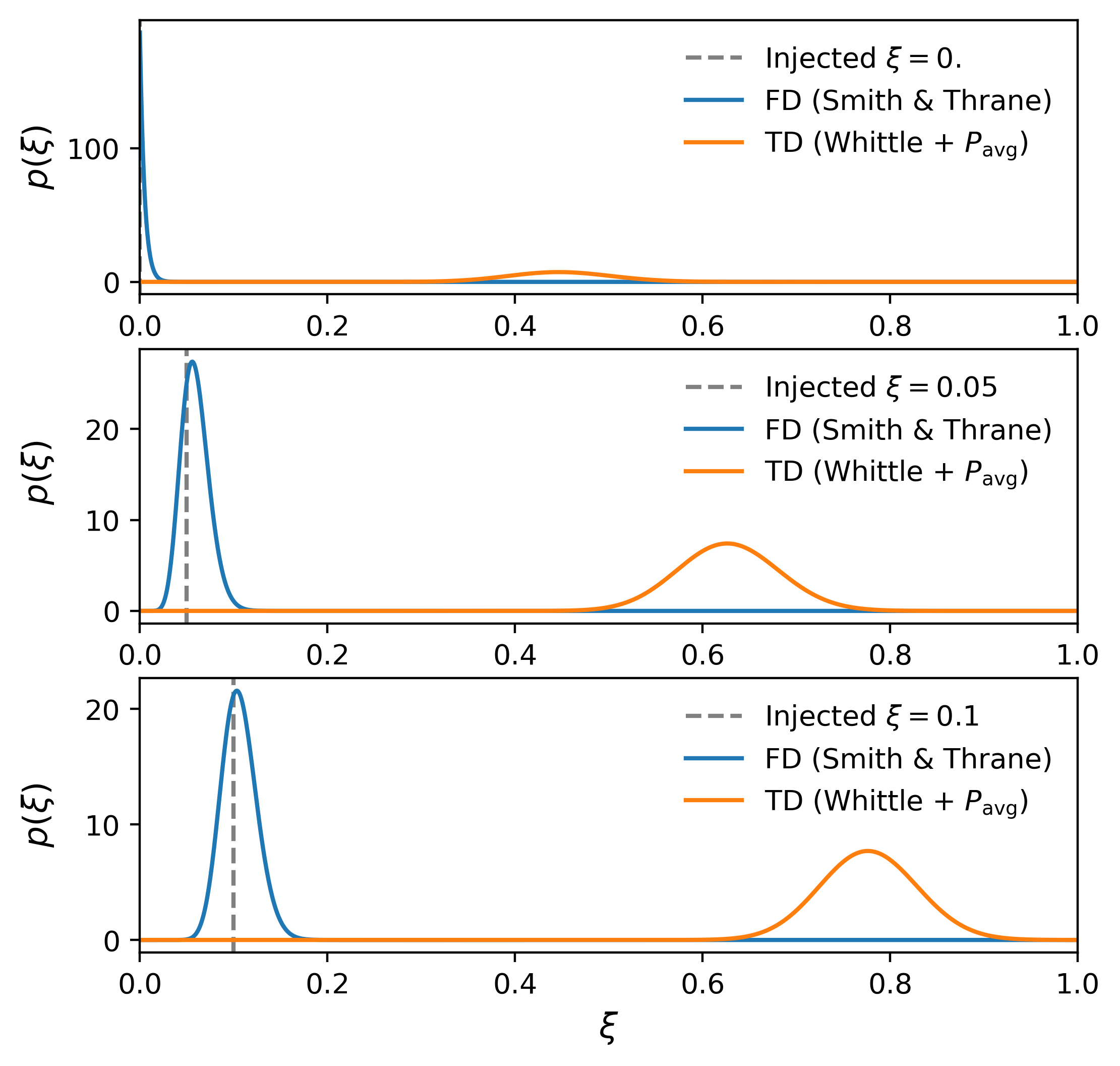}
    \caption{Comparison of analyzing frequency-domain and time-domain datasets. Each panel shows the posterior distribution $p(\xi\,|\,\mathcal{D})$, representing the inferred fraction of segments containing a signal, for different injected true values of $\xi$ indicated by the dashed vertical line. The frequency-domain (FD) results (blue curves) reproduce those of Smith \& Thrane \cite{Smith:2017vfk}, using simulated Gaussian noise with periodic boundary conditions and the true noise PSD to eliminate \textit{finite-duration} and \textit{noise uncertainty effects}. In contrast, the time-domain (TD) analysis (orange curves) incorporates these effects via the Whittle likelihood and an averaged PSD estimate. While the FD posteriors show excellent agreement with the injected $\xi$, the TD analysis exhibits bias due to incomplete modeling of finite-duration effects and noise uncertainty. Each analysis uses 3000 segments of duration $\unit[4]{s}$.}
    \label{fig:fd_true_psd_xi}
\end{figure}

For both FD and TD datasets, we perform parameter estimation on each segment following the method described in Section~\ref{subsec:PE}. For the FD datasets, we use the same (known) noise PSD in the Whittle likelihood (Eq. \ref{eq:whittle}) that we used to generate the data. For each segment, we compute the signal evidence \(Z_s^i\) and the noise evidence \(Z_n^i\) using Bilby~\cite{Ashton:2018jfp}, which employs nested sampling with the \texttt{dynesty} sampler~\cite{Speagle:2019ivv}. 
We do not explicitly marginalize over the time of coalescence, as our studies indicate that Bilby's~\cite{Ashton:2018jfp} marginalization method introduced a bias in the duty cycle at the time of this study.\footnote{We do not understand the source of this bias, and we are investigating it.}

In Fig.~\ref{fig:fd_true_psd_xi}, we present the posterior distribution \(p(\xi\,|\,\mathcal{D})\) for three injected duty cycle values. The analysis of the FD datasets (blue curves) accurately recovers the injected values in all cases, consistent with the findings of \cite{Smith:2017vfk}.
With the TD datasets, we again naively follow the procedure described in Section~\ref{subsec:PE}, but in the Whittle likelihood (Eq.~\ref{eq:whittle}), we do not use the same PSD used to generate the noise. Instead, we estimate a mean PSD, $P_{\rm avg}$ for each segment, as follows: we simulate an additional 128 seconds of noise, divide it into thirty-two $\unit[4]{s}$ sub-segments, apply windowing, and Fourier transform each segment. The resulting PSDs are then averaged to obtain $P_{\rm avg}$, which serves as the analysis PSD. These PSDs, along with the Fourier transform of the truncated TD datasets are used in the Whittle likelihood (without explicitly correcting for finite-duration effects and noise uncertainty). The results are shown in Fig.~\ref{fig:fd_true_psd_xi}.

In the following section, we introduce solutions to mitigate the systematic errors caused by finite-duration effects and noise uncertainty. \textit{Glitches} are not expected to arise in our simulated datasets and will therefore be addressed in a later section.

\section{Addressing systematic error}\label{sec:solution}
As discussed above, TD datasets are affected by two key sources of systematic error: uncertainty in the noise PSD and correlations between frequency bins due to finite-duration effects. Our ultimate goal is to address both effects simultaneously. 
We begin by tackling each source of error in isolation and then construct a strategy to account for them jointly.

To this end, we first present a solution for handling noise uncertainty using FD datasets, where finite-duration effects are absent. We then consider TD datasets in a simplified setting where the true PSD is assumed to be known, isolating the impact of finite-duration effects. Building on these two cases, the subsequent section introduces a method to approximately account for both systematics in combination.

\subsection{Noise Uncertainty}\label{subsec:psd_uncertainty}

In order to avoid bias from noise uncertainty, one must marginalize over uncertainty in the noise PSD:
\begin{equation}
\mathcal{L}_P(\tilde{d}\,|\, \theta, P_{\mathrm{avg}})=\int_0^{\infty} dP\,\mathcal{L}(\tilde{d}\,|\, \theta, P) \pi(P\,|\, P_{\mathrm{avg}}) .
\label{eq:marginalization}
\end{equation}
Here, $\mathcal{L}(\tilde{d}\,|\, \theta, P)$ is the (Whittle) likelihood of obtaining the data given model parameters $\theta$ and the true noise PSD $P$, and $\pi(P\,|\, P_{\mathrm{avg}})$ is our prior on the true PSD given the estimated PSD $P_{\mathrm{avg}}$ given by
\begin{equation}
\pi(P\,|\, P_{\mathrm{avg}}) \propto \mathcal{L}(P_{\mathrm{avg}}\,|\, P)\pi(P).
\label{eq:prior-P}
\end{equation}
Here, $\mathcal{L}(P_{\mathrm{avg}}\,|\, P)$ follows a $\chi^2$ distribution given the mean PSD $P_{\mathrm{avg}}$. 
We evaluate the integral in Eq.~\ref{eq:marginalization} assuming a uniform prior $\pi(P)$ to obtain the marginalized likelihood,
\begin{equation}
\begin{aligned}
\mathcal{L}_P(\tilde{d}\,|\,\theta, P_{\mathrm{avg}})&=\frac{2(N-1)}{\pi NT P_{\mathrm{avg}}}\left(1+\frac{2|\tilde{d} - \tilde{h}(\theta)|^2}{N T P_{\mathrm{avg}}}\right)^{-N} ,
\end{aligned}
\label{eq:fd_priors}
\end{equation}
where $N$ is the number of segments used to obtain the average PSD estimate. This problem has been thoroughly studied in the literature~\citep{Rover:2008yp,Banagiri:2019lon,student-t}, leading to slightly different solutions arising from variations in the underlying assumptions.

As shown in Appendix~\ref{appendix:psd}, marginalizing over the noise PSD uncertainty as in Eq. \ref{eq:fd_priors} removes the bias in parameter estimation that arises when the PSD is estimated from finite data. 
This approach effectively mitigates PSD-related bias in the frequency domain.

\subsection{Finite duration effects}\label{subsec:finite_duration}
As noted above, real GW data are produced as a time series that is divided into finite segments for analysis. 
As detailed in Ref.~\cite{windows}, Fourier transform of the finite-duration data segments results in correlations between frequency bins, described by a non-diagonal covariance matrix $C$ that is defined by the noise PSD and the windowing function applied in the time domain. The resulting likelihood is then given by
\begin{equation}
\mathcal{L}(\tilde{d} \mid \theta, C)=\frac{2}{T \operatorname{det} C} \exp \left[-\frac{2}{T}\langle\tilde{d}-\tilde{h}(\theta), \tilde{d}-\tilde{h}(\theta)\rangle_{{C}}\right]  ,
\label{eq:full_fd_likelihood}
\end{equation}
where we define the inner product
\begin{equation}
\langle\tilde{x}, \tilde{x}\rangle_{C}=\tilde{x}_i C_{ij}^{-1} \tilde{x}_j^* .
\label{eq:Cij}
\end{equation}
Ref.~\cite{windows} provides an algorithm for computing $C_{ij}$ and its inverse \textit{based on the known noise PSD} and the window function (see its Eqs. (15-16)).
In Fig.~\ref{fig:psd_acf_cij} (bottom), we reproduce the top-left panel of Fig.~12 from \cite{windows}, which shows the finite-duration covariance matrix within the frequency band $[12,800],\mathrm{Hz}$. This matrix is derived from the known PSD (top) using a Tukey window with parameter $\alpha = 0.1$, following the algorithm described in \cite{windows}. Off-diagonal elements in $\mathcal{C}$ are particularly prominent around frequency bins with sharp spectral features. Ref.~\cite{windows} also demonstrated that neglecting these off-diagonal elements can lead to a biased estimate of the duty cycle, whereas their proposed off-diagonal likelihood formulation proved effective in scenarios where the true PSD is assumed to be known (see their Fig. 9).

While the method used in this paper is in principle the same as proposed in \cite{windows}, 
we compute the likelihood in the time domain~\cite{Isi:2021iql,Miller:2023ncs,Wang:2024liy,Siegel:2024jqd} using the noise auto-correlation function (ACF), as part of optimizing the computational cost. 
Detailed discussions and comparisons with the frequency domain likelihood can be found in Appendix~\ref{appendix:finite-duration} and~\ref{appendix:td_fd_likelihood}.
This ACF-based time-domain likelihood is advantageous because, under the assumption of a rectangular window, the time-domain covariance matrix is Toeplitz, for which the matrix inversion can be performed in a computationally efficient way due to its special structure (see Appendix~\ref{appendix:finite-duration} for more details). \changed{This acceleration is crucial when we address both finite-duration effects and PSD uncertainty in Section \ref{subsec:both}.}
The middle panel of Fig.~\ref{fig:psd_acf_cij} shows the ACF corresponding to the PSD in the top panel.

The presence of off-diagonal elements makes the finite-duration likelihood relatively computationally expensive compared to the usual Whittle likelihood, which does not require matrix inversion.
We therefore employ importance sampling \citep{hom}, using the standard Whittle likelihood as a proposal distribution, and the finite-duration likelihood as a target distribution; see Appendix~\ref{appendix:reweighting} for details. As shown in Appendix~\ref{appendix:finite-duration}, marginalizing over the uncertainty in PSD estimation reduces the bias in duty cycle inference when compared to Fig.~\ref{fig:fd_true_psd_xi} (orange curves). However, an additional re-weighting procedure to correct for finite-duration effects is still necessary to achieve unbiased results when the known PSD is available.

\begin{figure*}[ht]
\includegraphics[width=\textwidth,height=0.423\textwidth]{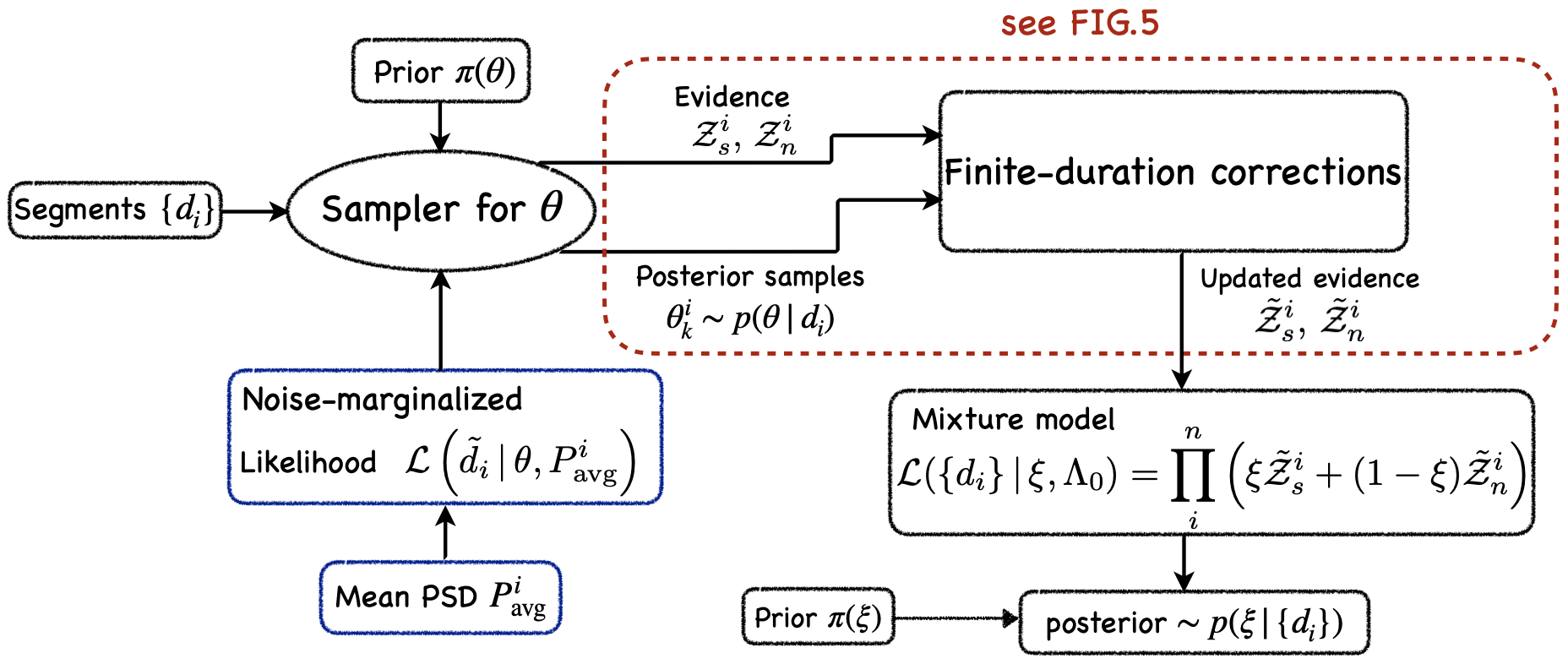}
\caption{The flowchart illustrating our analysis---similar to the Smith \& Thrane analysis shown in Fig.~\ref{fig:general_tbs} except we  take into account noise uncertainty and finite-duration effects (shown in colored cells). 
As above the data are divided into segments denoted $d_i$.
Each segment is analyzed to determine the binary black hole parameters $\theta$.
This time, however, we use a noise-marginalized, finite-duration likelihood described in the text (and depicted in more detail in Fig. \ref{fig:draw_tbs_pipeline}).
This step reweighs the two evidence values for each segment: ${\cal Z}_S$ for the signal hypothesis and ${\cal Z}_N$ for the noise hypothesis.
The evidence values are then input in the mixture likelihood, which is used to estimate the duty cycle $\xi$.
}
\label{fig:td_true_reweight}
\end{figure*}

\subsection{Simultaneously addressing both effects}\label{subsec:both}

In principle, one could marginalize over noise uncertainty using the finite-duration matrix, for example, by diagonalizing the matrix to obtain independent eigenmodes. 
However, it turns out that a computationally simpler approximation is adequate for our purposes here.
Instead of marginalizing over realizations of the noise PSD, we instead calculate an \textit{effective} PSD $\hat P$ that is implied by the marginalization procedure and use that to calculate the finite-duration covariance matrix.
In doing so, we ensure that our posterior for $\xi$ peaks in the correct place, though, it is not as wide as it would be if we actually carried out the marginalization.\footnote{In effect, we assume that the change in the posterior width is small compared to the width determined by the number of segments used in the calculation. When the number of segments is sufficiently large, the posterior width is underestimated, but it may be possible to correct for this in post-processing with a simple multiplicative factor.}

In order to obtain $\hat P$, we equate the noise-marginalized likelihood, which depends on the average noise PSD $P_\text{avg}$, with the original Whittle likelihood (given by the effective noise PSD $\hat P$) and solve for $\hat P$ in each frequency bin:
\begin{widetext}
\begin{equation}
\frac{2(N-1)}{\pi N T P_{\mathrm{avg}}}\left(1+\frac{2\left|\tilde{d}-\tilde{h}\left(f;\,{\theta^k}\right)\right|^2}{T N P_{\mathrm{avg}}}\right)^{-N}=\frac{2}{\pi T \hat{P}} \exp \left(-\frac{2\left|\tilde{d}-\tilde{h}\left(f;\,{\theta^k}\right)\right|^2}{T \hat{P}}\right) .
\label{eq:approximation}
\end{equation}
\end{widetext}

\begin{figure*}[t]
\includegraphics[width=0.8\textwidth,height=0.88\textwidth]{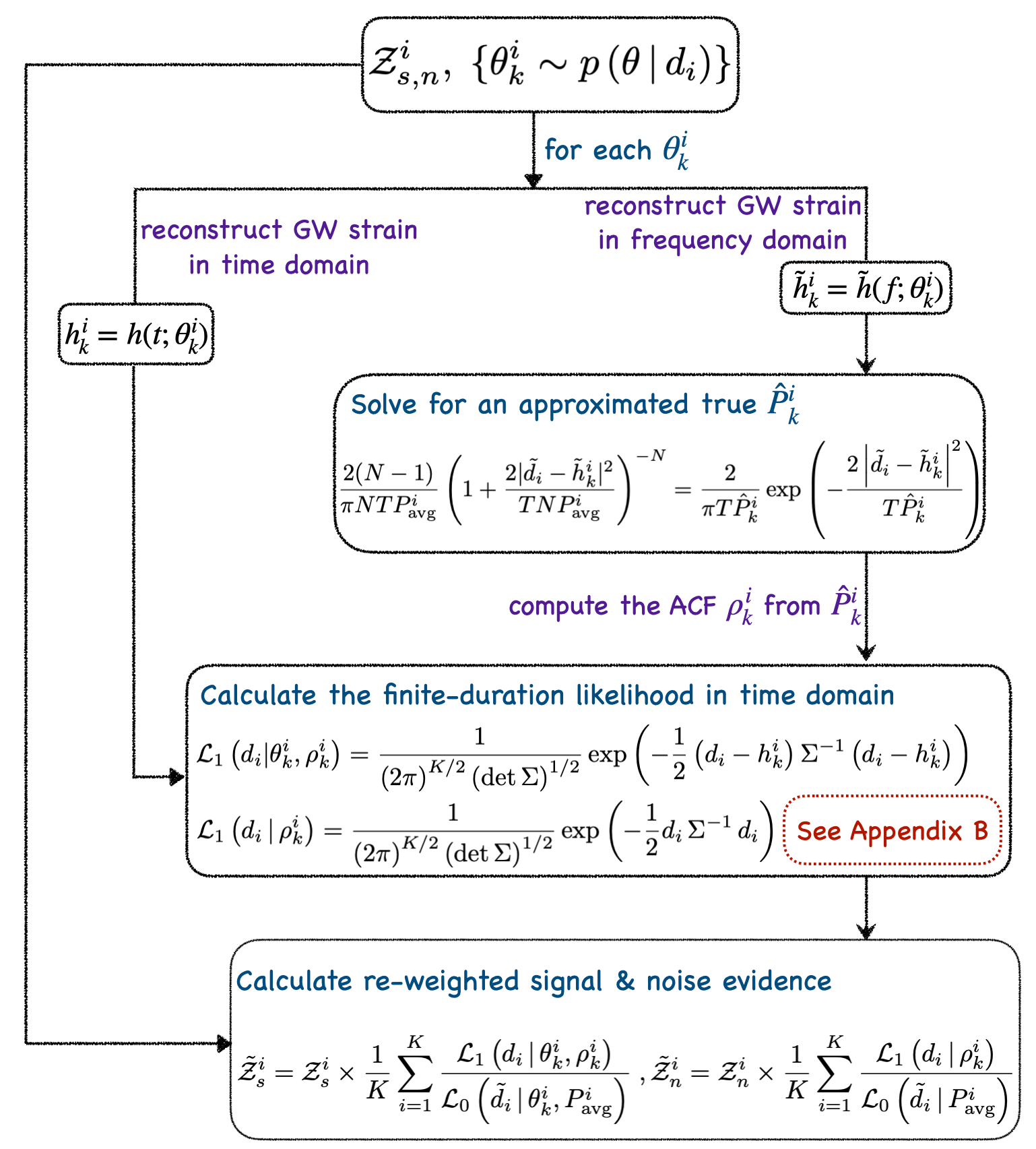}
\caption{Flowchart depicting the procedure for estimating $\hat{P}$, computing the finite-duration likelihoods for signal and noise, and reweighing the signal and noise evidences (further discussed in Section~\ref{subsec:both}). This procedure is applied to data from each segment.}
\label{fig:draw_tbs_pipeline}
\end{figure*}

We note that in the limit when $N \rightarrow \infty$, the two distributions in Eq. \ref{eq:approximation} asymptote each other, and $\hat{P}$ approaches the true noise PSD $P$. This calculation is carried for every posterior sample $\theta^k$. In other words, we get a unique $\hat{P}_k$ for each sample $\theta^k$, which we use for calculating the corresponding ACF$_k$. 
From the ACF, it is straightforward to compute the time-domain covariance matrix and hence the finite-duration likelihood in the time domain for both the signal and noise modelss. 
Once we have the updated likelihoods for all the posterior samples, we use them to obtain the reweighted evidence for the signal and noise model, following the standard reweighting procedure described in Appendix~\ref{appendix:reweighting}. These updated evidence values are then inserted into the mixture likelihood expression (Eq.~\ref{eq:total_mixture}) to estimate the duty cycle parameter. 
The procedure followed here is depicted in the flowcharts of Figs. \ref{fig:td_true_reweight} and~\ref{fig:draw_tbs_pipeline}. 

\begin{figure}[ht]
    \centering
    \includegraphics[width=1.0\linewidth]{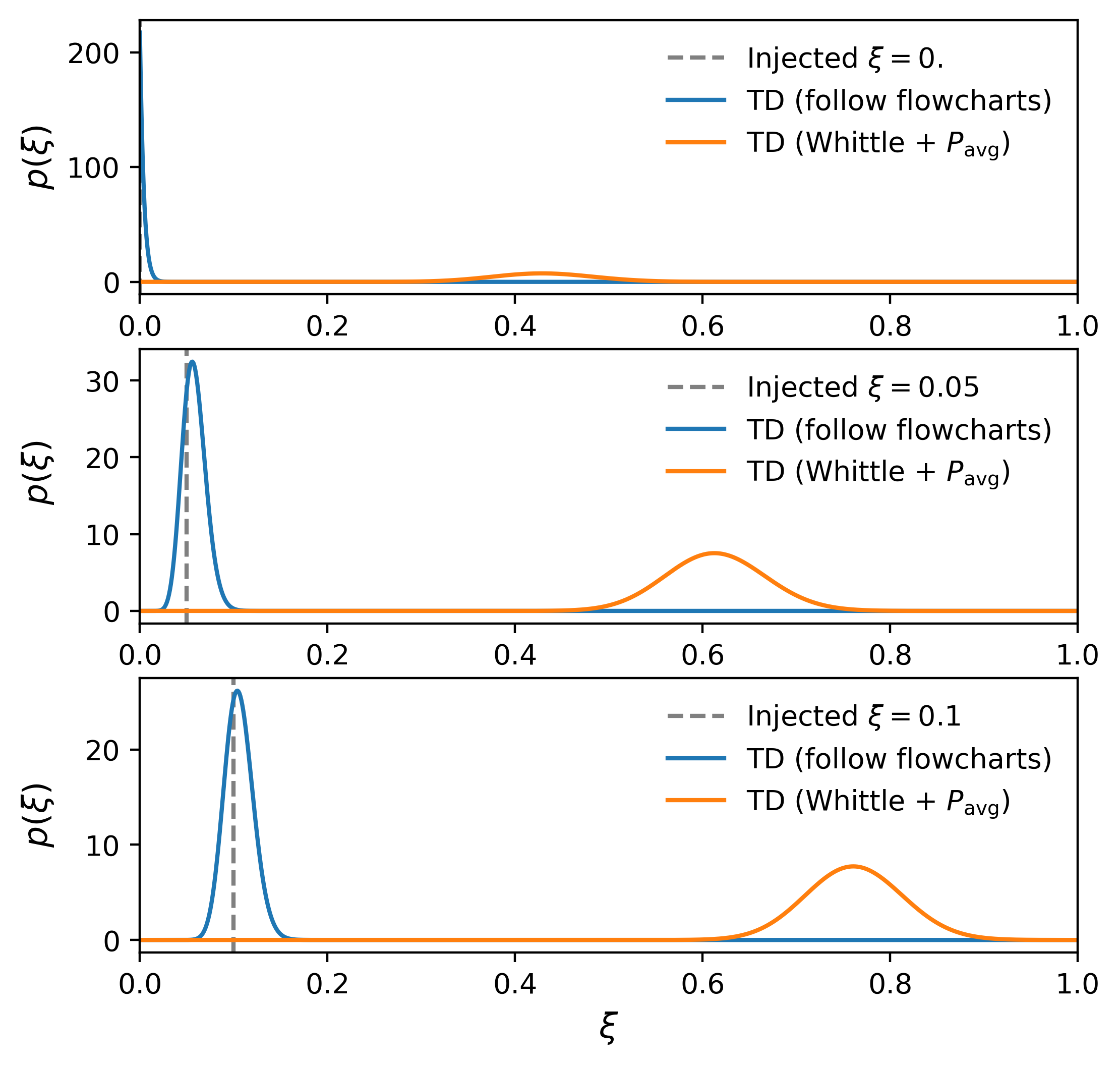}
    \caption{
        Removing both the noise uncertainty and finite-duration effects.
    Each panel shows the posterior distribution $p(\xi\,|\,\mathcal{D})$, which represents the fraction of segments containing a signal for different injected values of $\xi$.
    The posteriors (blue curves) align with the true values of $\xi$ by following the analysis steps outlined in the flowcharts (Figs.~\ref{fig:td_true_reweight} and~\ref{fig:draw_tbs_pipeline}), in contrast to the biased results (orange curves) where these effects are ignored.
    The plot is made using 3000 $\unit[4]{s}$ segments.
    }
    \label{fig:td_reweight_acf}
\end{figure}

Figure~\ref{fig:td_reweight_acf} shows the results of this formalism, applied to the same simulated data sets that resulted in biased $\xi$ posteriors in Section \ref{sec:demonstration} and shown in Fig. \ref{fig:fd_true_psd_xi}. The bias is successfully removed for all three simulated values of $\xi$, including the case where no signal was simulated ($\xi=0$).  While this method empirically works, we do not have any guarantees that we have an unbiased estimator of the target likelihood (the noise marginalized finite-duration likelihood) and hence remains an open problem for future. 

\subsection{Modeling  Glitches}
So far, we have shown how to simultaneously address the noise marginalization and finite-duration effects, provided that the noise is Gaussian. However, real data also includes non-Gaussian artifacts such as glitches. 
In this subsection, we describe our glitch model, adopted from Ref.~\cite{Smith:2017vfk}.

Assuming a two-detector network, we model glitches as binary black hole waveforms, with the distinction that the waveform in one detector is entirely uncorrelated with the waveform in the other. 
Of course, we do not believe that glitches look like binary black hole signals, but we can make this assumption in the name of conservatism: glitches that look like binary black hole signals represent the worst-case scenario for distinguishing the binary black hole background from noise.

We introduce parameters $\xi_g^{(1)}$ and $\xi_g^{(2)}$, which correspond to the glitch duty cycles in detectors one and two, respectively: the fraction of segments that contain a glitch. For each detector, we define single-detector glitch evidence:
\begin{equation}
\begin{aligned}
\mathcal{Z}_g^{(1)} & \equiv \int d \theta^{(1)} \mathcal{L}\left(d^{(1)} \mid \theta^{(1)}\right) \pi\left(\theta^{(1)}\right) \\
\mathcal{Z}_g^{(2)} & \equiv \int d \theta^{(2)} \mathcal{L}\left(d^{(2)} \mid \theta^{(2)}\right) \pi\left(\theta^{(2)}\right),
\end{aligned}
\end{equation}
where $\theta^{(1)}$ and $\theta^{(2)}$ are the signal parameters for (uncorrelated) glitches in detectors 1 and 2 respectively. The variables $d^{(1)}$ and $d^{(2)}$ are the strain data in each detector. We introduce $\mathcal{Z}_n^{(1)}$ and $\mathcal{Z}_n^{(2)}$ to denote the single-detector noise evidences. All of the evidences are computed following the procedure outlined in Figs. \ref{fig:td_true_reweight} and~\ref{fig:draw_tbs_pipeline}.
%
Following Ref.~\cite{Smith:2017vfk}, we ignore terms associated with simultaneous signal + glitch scenarios. Under this assumption, the resulting ``glitchy'' likelihood for each analyzed segment $d_i$ is:
\begin{equation}
\begin{aligned}
\mathcal{L}\left(d_i \mid \xi, \xi_g^{(1)}, \xi_g^{(2)}\right) \approx\; & \xi\left(1-\xi_g^{(1)}\right)\left(1-\xi_g^{(2)}\right) \mathcal{Z}_S+ \\
& (1-\xi)\left(1-\xi_g^{(1)}\right)\left(1-\xi_g^{(2)}\right) \mathcal{Z}_N+ \\
& (1-\xi) \xi_g^{(1)}\left(1-\xi_g^{(2)}\right) \mathcal{Z}_g^{(1)} \mathcal{Z}_N^{(2)}+ \\
& (1-\xi)\left(1-\xi_g^{(1)}\right) \xi_g^{(2)} \mathcal{Z}_N^{(1)} \mathcal{Z}_g^{(2)}+ \\
& (1-\xi) \xi_g^{(1)} \xi_g^{(2)} \mathcal{Z}_g^{(1)} \mathcal{Z}_g^{(2)}.
\end{aligned}
\label{eq:glitch_likelihood}
\end{equation}

\section{Application to time-reversed LIGO O3b data}\label{sec:real-o3-noise}

We select 3,000 non-overlapping 4\,s segments from the LIGO O3b observation run, which spanned November 1, 2019, to March 27, 2020. The data are taken from the \texttt{DCS-CALIB\_STRAIN\_CLEAN\_SUB60HZ\_C01} frames during science-quality times, as indicated by the \texttt{DMT-ANALYSIS\_READY:1} data-quality flag. Segments containing prominent glitches, as identified in~\cite{ZweizigRiles2020}, are excluded. \changed{Additionally, we imposed a criterion that each chunk of data must exceed 200\,s in length for both detectors. This ensures that for every $\unit[4]{s}$ segment in that chunk, there is at least 128\,s of neighboring data available for estimating the PSD.} As a result, our analysis is limited to segments\footnote{\changed{Our final selection of 3000 4\,s segments, was drawn from the continuous observation period between November 15, 2019 and November 19, 2019 while ensuring that all aforementioned data conditions were satisfied.}} where science-quality data are available from both LIGO detectors and no known glitches are present. Nevertheless, weak, unclassified glitches may still remain and potentially contaminate our inferences.

To test the performance of our algorithm in realistic noise conditions, we require data that resemble detector noise but are free of gravitational-wave signals. Since the LIGO detectors cannot be shielded from astrophysical signals, we address this challenge by time-reversing the data. This approach preserves many aspects of the non-stationary noise features while suppressing gravitational-wave signals, which no longer match our forward-modeled templates.

For each \(4\,\mathrm{s}\) segment, we estimate the PSD using \(128\,\mathrm{s}\) of neighboring data (32 segments before and after). This PSD is then used for estimation of BBH parameters.
We perform two separate analyses on these 3000 time-reversed segments. The first uses the mixture likelihood defined in Eq.~\ref{eq:simple_mixture} to estimate the duty cycle parameter \(\xi\), which does not account for glitches but it does account for biases due to the noise PSD uncertainty and due to finite-duration effects. The second employs the glitch-modeling likelihood described in Eq.~\ref{eq:glitch_likelihood}, explicitly accounting for glitches as well as for the biases due to the noise PSD uncertainty and finite-duration effects.

\begin{figure}[ht]
    \centering
    \includegraphics[width=1.0\linewidth]{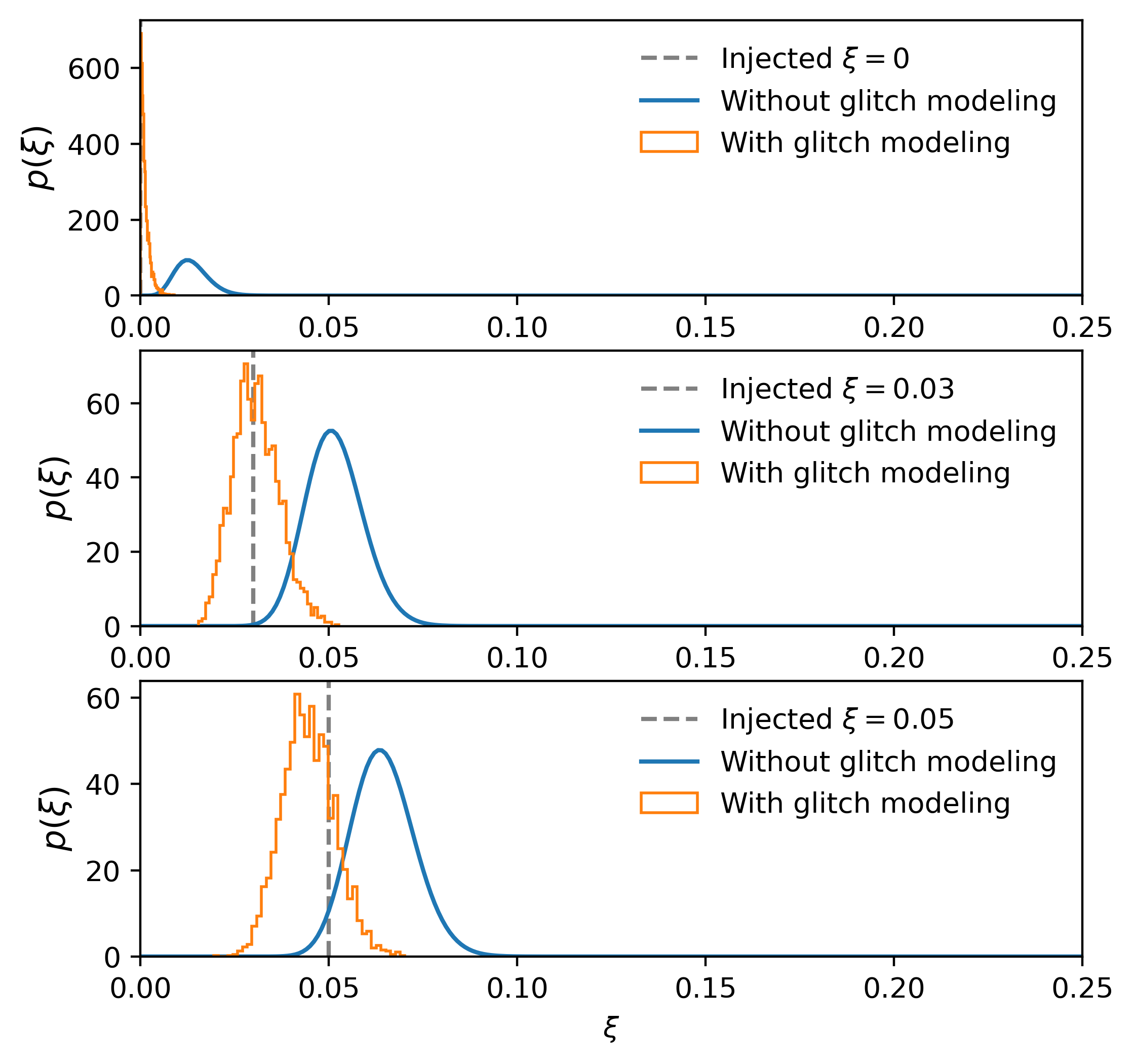}
    \caption{Posterior distributions of the duty cycle parameter, $p(\xi\,|\,\mathcal{D})$, constructed using time-reversed real data from two LIGO detectors during the O3b observation run. The two colors represent analyses with (orange) and without (blue) the inclusion of glitch modeling. The dashed vertical lines in the three panels indicate three different injected values of $\xi$, with the top panel corresponding to the case with no injection. The analysis is performed on 3000 non-overlapping segments, each lasting 4 seconds, where each segment is analyzed using our modified pipeline illustrated in Figs.~\ref{fig:td_true_reweight} and~\ref{fig:draw_tbs_pipeline}. Each segment is analyzed coherently using data from both detectors (H1L1), as well as independently using data from each detector (H1 and L1). The blue curve (without glitch modeling) shows the result from the H1L1 coherent analysis, while the orange curve (with glitch modeling) incorporates both the H1L1 coherent and the single-detector (H1 and L1) analyses, as described in Eq.~\ref{eq:glitch_likelihood}. These plots demonstrate that modeling glitches enables us to accurately recover the true duty cycle values.
    }
    \label{fig:real_H1L1}
\end{figure}

Figure~\ref{fig:real_H1L1} shows the posterior distributions for the duty cycle \(\xi\) from both analyses. In the no-glitch model (blue curves), the analysis uses only the coherent combination of the two detectors (H1L1). In contrast, the glitch-modeling analysis (orange curves) supplements the coherent term with incoherent single-detector  contributions (H1 and L1) in order to decouple the glitch contamination from the duty cycle, following Eq.~\ref{eq:glitch_likelihood}.

The three panels correspond to three different injected values of \(\xi\), marked by dashed vertical lines, with the top panel showing results for no injection. Across all cases, the inclusion of glitch modeling improves the recovery of the true duty cycle values. This demonstrates the importance of modeling glitches to achieve robust duty cycle estimation in real detector data. 
Thus, our analysis of time-reversed data shows that LIGO noise is not well characterized as Gaussian, but that we can avoid false positives using the glitchy likelihood in Eq.~\ref{eq:glitch_likelihood}.\footnote{Astute readers may wonder why Smith \& Thrane~\cite{Smith:2017vfk} were able to obtain unbiased estimates of $\xi$ from time-reversed data without accounting for finite-duration effects or PSD marginalization. While the exact reason remains unclear, our analysis shows that in Gaussian noise, neglecting these effects leads to biased inferences. It is possible that in~\cite{Smith:2017vfk}, the glitch likelihood absorbed the systematic errors arising from finite-duration effects and PSD uncertainty during their time-reversed analysis.}

\section{Discussion and Summary}\label{sec:summary}
The gravitational wave events detected so far represent only a tiny fraction of the ${\cal O}(10^5)$ binary black hole mergers that take place every year. 
In this paper, we implement the search for sub-threshold binary black holes proposed by Smith \& Thrane~\cite{Smith:2017vfk}, while addressing a number of subtleties that make this analysis difficult to carry out in practice.
In particular, we address effects arising from uncertainty in the noise model and the finite duration of analysis segments.
We show that these innovations are essential for analyzing real data.
After taking these effects into account, we show it is possible to analyze time-reversed LIGO data without detecting a false-positive signal.
In doing so, we clear the path for an optimal search for sub-threshold binary black hole signals.


There are a number of areas for future improvement.
Observations of binary black holes from clearly resolved gravitational-wave events~\cite{LIGOScientific:2021djp} have revealed various insights into the masses, spins, and merger rates of merging binary black holes~\cite{KAGRA:2021duu}.
In this paper, we do not address prior mismatch that occurs when our model for the distribution of binary black hole parameters (e.g., masses and spins) does not reflect the actual distribution of binary black holes in the Universe.
However, it should be possible to extend our analysis to allow for a more flexible population model as in Ref.~\cite{Smith:2020lkj}.

Ideally, our analysis should be carried out with as much data as possible. 
The computational cost of a phase-coherent analysis remains challenging.
Smith \& Thrane~\cite{Smith:2017vfk} calculate that a single day of design-sensitivity LIGO data is probably adequate to detect a signal from subthreshold binary black holes. Since LIGO has not yet reached design sensitivity, more than one day would likely be required---perhaps $\approx 10$ days.
Our analysis of 3 hours and 20 minutes of LIGO data required $\sim 8000$ core hours on contemporary hardware. 
We therefore estimate it would take $\sim 5\times10^5$  core hours of time to analyze ten days of LIGO's most sensitive data. Being massively parallelizable, on a dedicated 1000-core cluster, this is  equivalent to less than a month of wall-time.

We anticipate that many of the current computational challenges can be mitigated with the integration of advanced techniques into various stages of the analysis. The most computationally expensive component is running stochastic samplers to perform BBH parameter estimation on each 4\,s data segment. To reduce processing time, methods to expedite likelihood calculations, including reduced order quadrature~\cite{ROQ,Qi:2020lfr}, relative binning~\cite{Zackay:2018qdy,Cornish:2021lje,Krishna:2023bug}, and explicit marginalization over parameters~\cite{intro,Lange:2018pyp,Roulet:2024hwz} have been developed. Recent advances in fast ML-based methods, including likelihood-free inference for learning posterior distributions~\cite{DINGO,AMPLFI} and the use of normalizing flows as a proposal distribution to accelerate samplers~\cite{NESSAI,JIM,Nautilus,pocomc}, offer promising alternatives. While these techniques have primarily been employed on loud BBH signals, they are likely extendable to the analysis of sub-threshold signals with appropriate adjustments.  Additionally, tasks such as likelihood reweighting for finite-duration corrections present further opportunities for significant acceleration using ML-based approaches. 
Ultimately, this may enable us to analyze the entire LVK dataset using currently available computational resources.

\begin{acknowledgments}
\changed{The authors thank Sharan Banagiri, Sylvia Biscoveanu, Rory Smith, Jessica Lawrence, Kevin Turbang, Joseph Romano, Tom Callister, Zoheyr Doctor, Gregory Ashton, Michael Coughlin, Nick Van Remortel, Haowen Zhong, Thomas Dent, and Suvodeep Mukherjee for valuable discussions and comments. We are especially grateful to Arianna Renzini for conducting the internal collaboration review and for providing detailed feedback that helped improve the quality of the manuscript}.
We are grateful
for computational resources provided by the LIGO Laboratory and supported by National Science Foundation
Grants PHY-0757058 and PHY-0823459.
X.K.\,  M.S.\ and V.M.\ acknowledge the NSF support under grant numbers PHY-2409173 and PHY-2110238.  X.K. was partially supported by the University of Minnesota Data Science Initiative with funding made available by the MnDrive initiative.
M.S.\ acknowledges the support from Weinberg Institute for Theoretical Physics at the University of Texas at Austin.
E.T.\ is supported through Australian Research Council (ARC) Centres of Excellence CE170100004, CE230100016, Discovery Projects DP230103088 \& DP250100373, and LIEF Project LE210100002.
This material is based upon work supported by NSF's LIGO Laboratory which is a major facility fully funded by the National Science Foundation.
\end{acknowledgments}

\appendix
\section{Derivation and comparison of the marginalized likelihood over PSD uncertainties}
\label{appendix:psd}
The Whittle likelihood for the noise $\tilde{n}_i$ at given frequency bin $f_i$ can be expressed as
\begin{equation}
\mathcal{L}_{\mathrm{Whittle}}\left(\tilde{n}_i\,|\,P_i\right)=\frac{2}{\pi T P_i}\exp \left(-\frac{2|\tilde{n}_i|^2}{T P_i}\right),
\end{equation}
\changed{where $P_i$ represents the true PSD evaluated at the same frequency bin $f_i$.} The real and imaginary part of the noise in the frequency domain are characterized by a Gaussian distribution $N(0,\frac{T}{4}P_i)$. 

Assuming we have $N$ randomly generated noise segments, the sample mean of the PSD is given by $P_{\mathrm{avg},\,i}=\frac{2}{NT}\sum_{k=1}^{N}\tilde{n}_k(f_i) \tilde{n}_k^*\left(f_i\right)$\footnote{\changed{Here, the index $k$ represents the $k$th segment in N random noise segments.}}. The distribution of $P_{\mathrm{avg},i}$ follows\footnote{The variable $2NP_{\mathrm{avg},\,i}/P_i$ follows a chi-square distribution $\chi^2\left(2N\right)$, with $2N$ degrees of freedom.} the relation
\begin{equation}
p(P_{\mathrm{avg},\,i}\,|\,N,P_i) = \frac{N^N}{\Gamma\left(N\right)P_i^N}\left(P_{\mathrm{avg},\,i}\right)^{N-1}\exp\left(-\frac{NP_{\mathrm{avg},\,i}}{P_i}\right).
\end{equation}
Bayes’ Theorem defines the posterior distribution of true PSD $\pi(P_i\,|\, N, P_{\mathrm{avg},\,i})$ given the estimated one $P_{\mathrm{avg},\,i}$ as follows
\begin{equation}
\pi(P_i\,|\, N, P_{\mathrm{avg},\,i}) \propto p(P_{\mathrm{avg},\,i}\,|\, N, P_i)\pi(P_i).
\end{equation} 
Assuming uniform prior on $P_i$, the normalized posterior distribution $\pi(P_i\,|\,N, P_{\mathrm{avg},\,i})$ can be obtained\footnote{The integral formula $\int z^N\exp(-kz)dz = \frac{1}{k^{N+1}}\Gamma(N+1)$ has been used here.} as:
\begin{equation}
\begin{aligned}
\pi(P_i\,|\,N, P_{\mathrm{avg},\,i}) =& \frac{N^{N-1}(N-1)}{\Gamma(N) P_i^N} \exp \left(-\frac{N P_{\mathrm{avg},\,i}}{P_i}\right) \\
&\times \left(P_{\mathrm{avg},\,i}\right)^{N-1}.
\end{aligned}
\end{equation}
Then, the likelihood marginalized over PSD uncertainty is expressed as:
\begin{equation}
\begin{aligned}
\mathcal{L}\left(\tilde{n}_i\,|\,N,P_{\mathrm{avg},\,i}\right) &= \int \mathcal{L}_{\mathrm{Whittle}}\left(\tilde{n}_i\,|\,P_i\right) \pi(P_i\,|\, N, P_{\mathrm{avg},\,i}) dP_i\\
&= \frac{2\left(N-1\right)}{N\pi T P_{\mathrm{avg},\,i}}\left(1+\frac{2|\tilde{n}_i|^2}{NTP_{\mathrm{avg},\,i}}\right)^{-N}.
\end{aligned}
\label{eq:Uniform_likelihood}
\end{equation}
In Ref.~\cite{Rover:2008yp}, the prior $\pi(P_i)$ is modeled as a scaled inverse $\chi^2$ distribution with scale parameter $P_{\mathrm{avg},\,i}$ and degrees of freedom $2N$, resulting in a slightly different form of the marginalized likelihood compared to Eq.~\ref{eq:Uniform_likelihood}. 
\changed{Although we do not present the details in this work, supporting studies carried out separately indicate that the uniform prior adopted here provides a more accurate recovery of the duty cycle than the one used in \cite{Rover:2008yp}}.

\begin{figure}[t]
    \centering
    \includegraphics[width=1.0\linewidth]{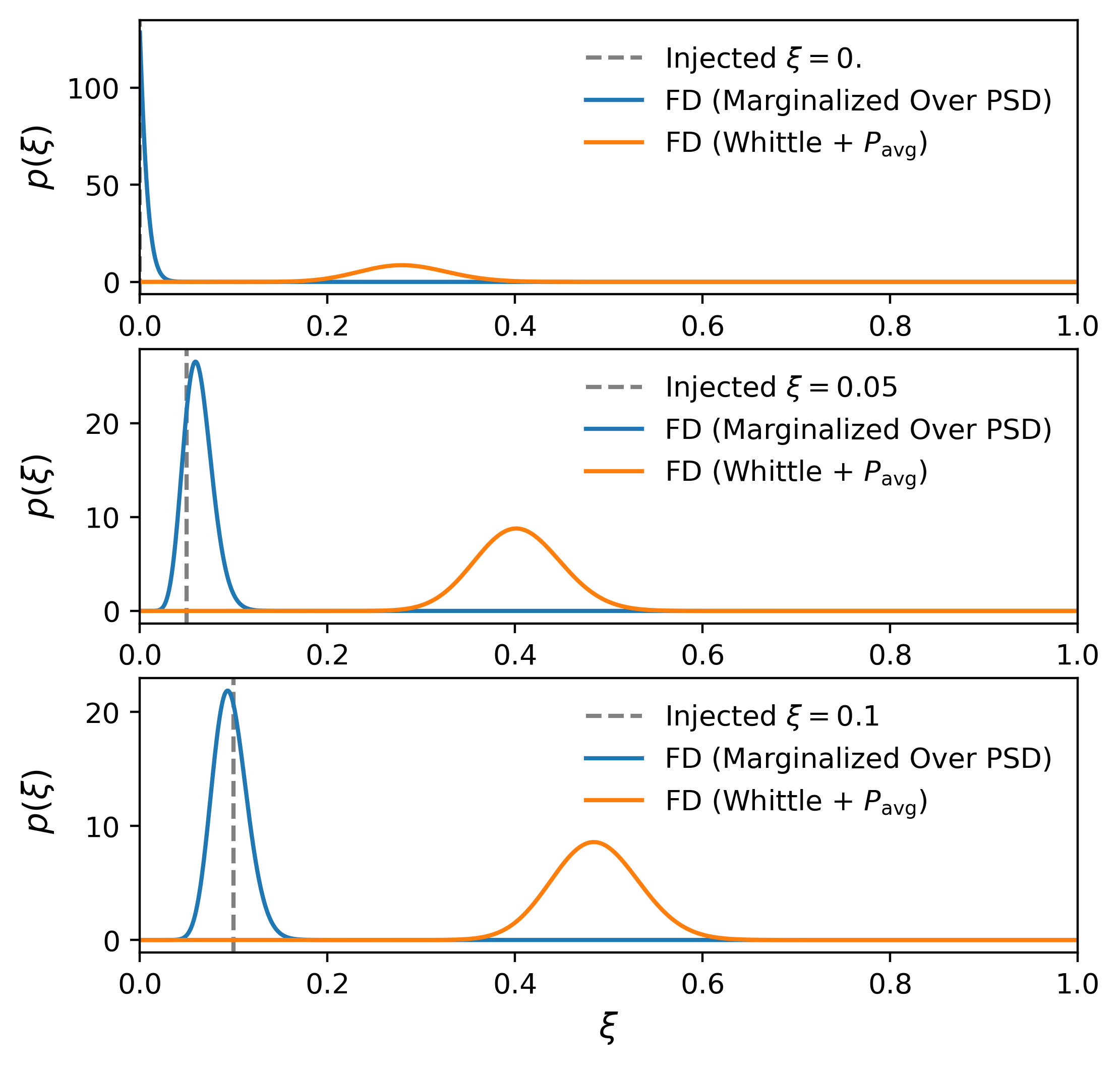}
    \caption{Removing the PSD uncertainty effects.
    Each panel shows the posterior distribution $p(\xi\,|\,\mathcal{D})$ for a different injected value of $\xi$, the fraction of segments containing a signal. Gaussian noise is simulated in the frequency domain (FD) so that there are no finite-duration effects. We employ the Whittle likelihood with the mean-averaged PSD (without modeling PSD uncertainty, orange curve), and the marginalized likelihood (Eq.~\ref{eq:Uniform_likelihood}, blue curve) to perform parameter estimation. The plot is made using 3000 $\unit[4]{s}$ segments.
    }
    \label{fig:fd_marg_compare_xi}
\end{figure}

To demonstrate the utility of the marginalized likelihood derived above, we construct controlled simulations in the frequency domain. While in real gravitational-wave analyses the data is collected in the time domain and transformed via Fourier transform (introducing windowing and finite-duration artifacts), our goal here is to isolate the impact of PSD uncertainty. By working directly in the frequency domain, we avoid complications arising from time-domain windowing and finite-length effects, allowing for a cleaner comparison between the standard and marginalized likelihoods.

Specifically, we generate both noise and signal frequency series where the noise is drawn from a standard normal distribution and scaled by the square root of the true PSD (shown in Fig.~\ref{fig:psd_acf_cij}\,(top)) at each frequency bin. For each analyzed 4 s segment, an independent 128 s noise realization is generated and used to estimate the mean PSD $P_{\mathrm{avg}}$, via averaging over thirty-two 4 s segments with matching frequency resolution. This ensures that the PSD used in the likelihood function is obtained in a consistent way with standard analysis practice, but without involving Fourier transforms or encountering leakage and windowing issues.

In Fig.~\ref{fig:fd_marg_compare_xi}, we present the posterior distribution $p(\xi\,|\,\mathcal{D})$ for three different injected duty cycle values. We compare results from the standard Whittle likelihood using the mean PSD (without modeling PSD uncertainty, yellow curves) to those from the marginalized likelihood (Eq.~\ref{eq:Uniform_likelihood}, blue curves). The marginalized likelihood successfully reduces the bias introduced due to uncertain PSD estimates.

\section{Finite duration likelihood}
\label{appendix:finite-duration}
A gravitational wave detector measures a discrete time series data stream $\mathbf{d}$ that contains potential signal $\mathbf{s(\theta)}$ contaminated by additive, Gaussian noise $\mathbf{n}$. The Gaussian noise can be described by a multivariate normal distribution
\begin{equation}
\mathbf{n} \sim \mathcal{N}(0, \boldsymbol{\Sigma})
\end{equation}
with mean $0$ and time-domain covariance matrix $\boldsymbol{\Sigma}$. The distribution of the noise implies that the likelihood function (i.e., the distribution of data $\mathbf{d}$ conditioned on a signal $\mathbf{s(\theta)}$) is
\begin{equation}
\mathcal{L}(\mathbf{d}\,|\, \theta,\,\boldsymbol{\Sigma})= \frac{\exp\left(-\frac{1}{2}(\mathbf{d}-\mathbf{s(\theta)})^T \boldsymbol{\Sigma}^{-1}(\mathbf{d}-\mathbf{s(\theta)})\right)}{\left(2\pi\right)^{K/2}\left(\operatorname{det} \boldsymbol{\Sigma}\right)^{1/2}},
\label{eq:td_likelihood}
\end{equation}
where $K$ is the total number of samples.

For a stationary Gaussian process\footnote{This assumption is valid for real data collected over several minutes.}, the covariance takes a particularly simple (symmetric Toeplitz) form
\begin{equation}
\Sigma_{ij}=\rho(|i-j|),
\label{eq:acf}
\end{equation}
where $\rho(k)$ is the auto-correlation function (ACF) and $i, j$ represent the discrete time indices of the time series. If, in addition to stationarity, we impose periodic boundary conditions, then $\rho(m)=\rho(K-m)$ and $\Sigma_{ij}$ will be circulant. Circulant matrices are diagonalized by the discrete Fourier transform~\cite{1984Unser}. The noise Fourier amplitudes become independent random variables with variance described by the PSD as a function of frequency, $P(f)$, which gives the Whittle likelihood that has been widely used in GW community.

However, it is important to note that the Whittle likelihood is merely an approximation of the true multivariate Gaussian distribution~\cite{1984Unser,Rover:2008yp}. Additionally, conventional gravitational wave analyses that utilize the Whittle likelihood depend on selecting segments of time-domain detector data in such a way that the signal is safely distanced from the edges of the segments. These edges are tapered—typically using a Tukey window function—to ensure a smooth transition and prevent leakage during the Fourier transformation. In the frequency domain, the windows\footnote{We also need to recognize that any window functions used to taper the edges of the segments can partially compromise the Gaussian property.} play a crucial role in transforming covariances between neighboring time bins into frequency bins. However, when the PSD is available, we can also efficiently compute the likelihood in the time domain. The Wiener-Khinchin theorem establishes a direct relationship between the auto-correlation function and the PSD via the Fourier transform. The computation of the time-domain term $(\mathbf{d}-\mathbf{s(\theta)})^T \boldsymbol{\Sigma}^{-1}(\mathbf{d}-\mathbf{s(\theta)})$ in Eq.~\ref{eq:td_likelihood} can be optimized by taking the gravitational wave signal in the time domain and utilizing the \texttt{solve\_toeplitz} function available in the SciPy package. The significant improvement\footnote{The \texttt{solve\_toeplitz} function efficiently combines the computation of the covariance inverse and the multiplication of the inverse by a vector into one step by fully utilizing properties of the Toepolitz matrix. It is important to note that while the computation speed with \texttt{solve\_toeplitz} is much faster than directly computing the multiplication of the covariance inverse and a vector (\changed{with a speedup proportional to the dimensionality of the matrix}), it is still about 10 times slower than evaluating the Whittle likelihood. This limitation prevents us from utilizing this approach for parameter estimation directly in the time domain.} in the speed of finite-duration likelihood calculation in the time domain will be crucial for the numerical approach introduced in the Section~\ref{subsec:both}, where we addressed both effects simultaneously via a final re-weighting step.

\begin{figure}
    \centering
    \includegraphics[width=1.0\linewidth]{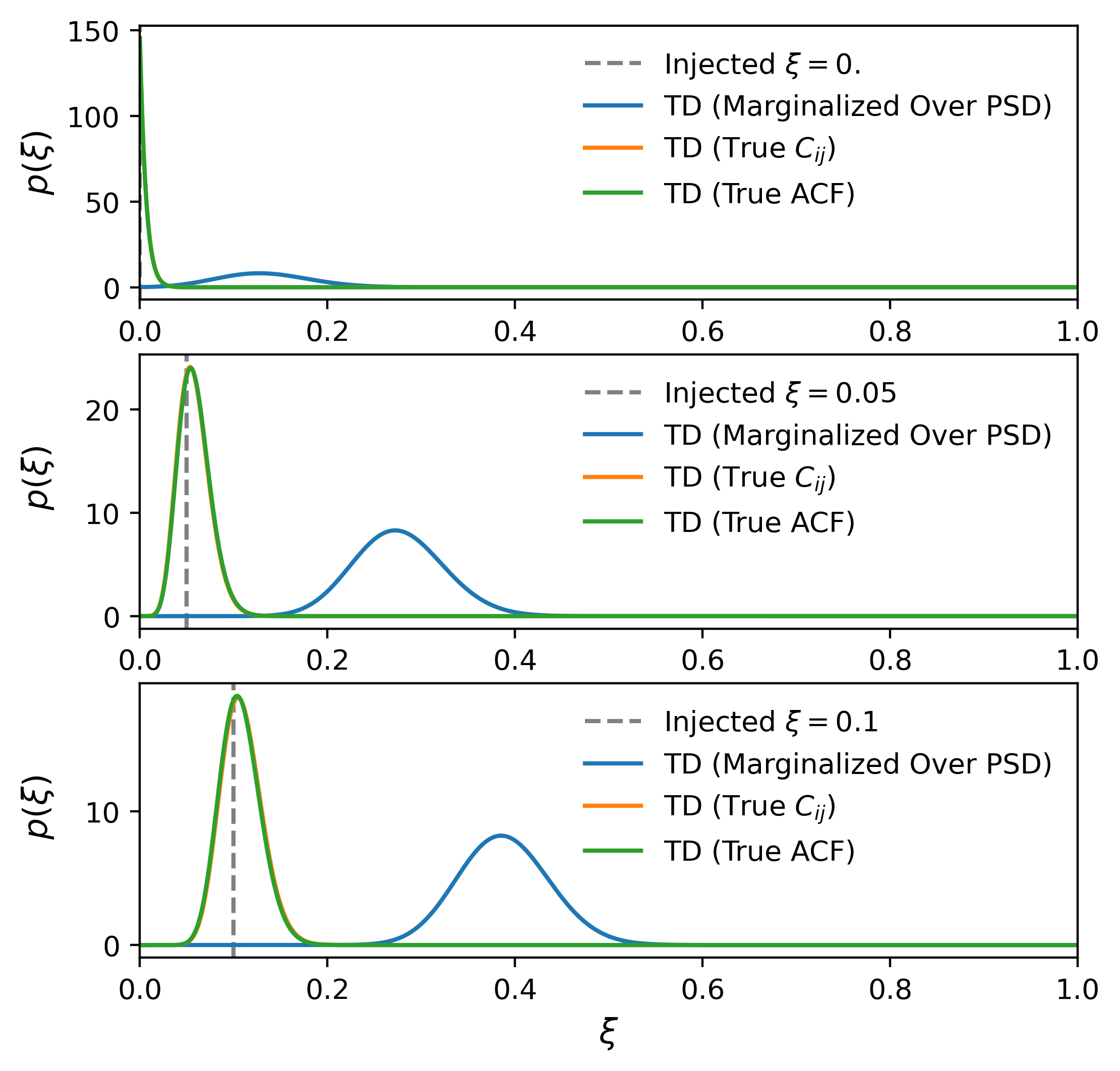}
    \caption{Removing the finite duration effects.
    Each panel shows the posterior distribution $p(\xi\,|\,\mathcal{D})$, which represents the fraction of segments containing a signal. 
    Each panel shows a different true value for $\xi$ indicated by the dashed vertical line.
    Gaussian noise and signal datasets are simulated in the time domain (TD). We employ the marginalized likelihood (Eq.~\ref{eq:Uniform_likelihood}) for the initial parameter estimation (blue curves). Subsequently, we re-weight the signal and noise evidence by employing the $C_{ij}$ (orange curves) and the ACF (green curves), computed using the true (known) PSD, into evaluating the updated finite-duration likelihood. The plot is made using 3000 $\unit[4]{s}$ segments.
    }
    \label{fig:td_true_compare_xi}
\end{figure}

\section{Likelihood and Evidence Re-weighting}
\label{appendix:reweighting}
A crucial aspect of our analysis when analyzing time-domain data is the acquisition of Bayesian evidence and posterior samples that account for finite-duration effects \textit{via} a full likelihood that  models the data with a non-diagonal covariance matrix. However, directly conducting nested sampling on the full (target) likelihood can be computationally intensive, while the approach based on the Whittle (proposal) likelihood can be much faster although less accurate. 

By conducting Bayesian inference with the proposal likelihood (denoted by the subscript 0), we obtain the "proposal posterior samples" for the distribution
\begin{equation}
p_{0}(\theta\,|\, d)=\frac{\mathcal{L}_{0}(d\,|\, \theta) \pi(\theta)}{\mathcal{Z}_{0}},
\end{equation}
along with its Bayesian evidence $\mathcal{Z}_0$. Our objective is to derive expressions for the target posterior (denoted by the subscript 1):
\begin{equation}
p_1(\theta\,|\, d)=\frac{\mathcal{L}_1(d\,|\, \theta) \pi(\theta)}{\mathcal{Z}_1},
\end{equation}
and the target Bayesian evidence
\begin{equation}
\mathcal{Z}_1=\int d \theta \mathcal{L}_1(d\,|\, \theta) \pi(\theta)
\end{equation}
expressed in terms of the proposal likelihood and its evidence.

The proposal quantities are linked to the target quantities through a weight factor. By multiplying the target posterior by unity, we can derive
\begin{equation}
\begin{aligned}
p_1(\theta\,|\, d) & =\frac{\mathcal{L}_{0}(d\,|\, \theta)}{\mathcal{L}_{0}(d\,|\, \theta)} \frac{\mathcal{L}_1(d\,|\, \theta) \pi(\theta)}{\mathcal{Z}_1} \\
& =w(d\,|\, \theta) \frac{\mathcal{L}_{0}(d\,|\, \theta) \pi(\theta)}{\mathcal{Z}_1}
\end{aligned}
\end{equation}
where $w(d\,|\, \theta) \equiv \mathcal{L}_1(d\,|\, \theta)/\mathcal{L}_{0}(d\,|\, \theta)$ represents the weight function. By multiplying by unity once more, we arrive at the following expression for the evidence:
\begin{equation}
\begin{aligned}
\mathcal{Z}_1 & =\mathcal{Z}_{0} \int d \theta\, p_{0}(\theta\,|\, d)\left(\frac{\mathcal{L}_1(d\,|\, \theta)}{\mathcal{L}_{0}(d\,|\, \theta)}\right) \\
& =\frac{\mathcal{Z}_{0}}{N} \sum_{i=1}^{N}w\left(d\,|\, \theta_i\right) .
\end{aligned}
\end{equation}
In the second line, the integral has been replaced with a discrete sum over $N$ proposal posterior samples.

\section{\changed{Time and Frequency domain likelihoods, and the Whittle approximation}}
\label{appendix:td_fd_likelihood}

\begin{figure}[t]
    \centering
\includegraphics[width=1.0\linewidth]{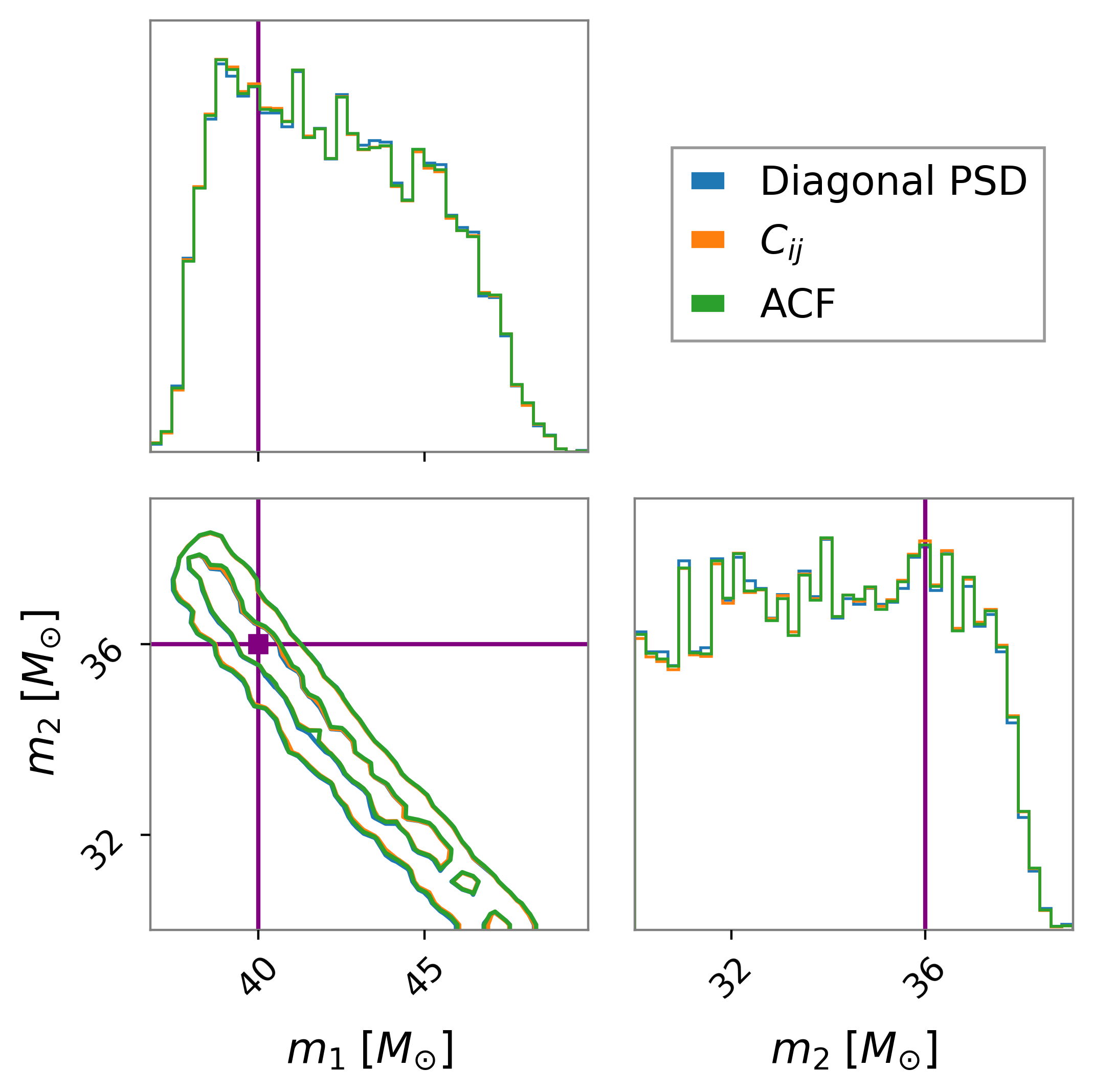}
    \caption{Intrinsic parameters posterior distributions for the mock data analysis using the diagonal Whittle likelihood (blue) and the full likelihood evaluated in the time (green, ACF) and frequency (orange, $C_{ij}$) domain. The primary and secondary mass $(m_1, m_2)$ refer, respectively, to the more-massive and less-massive component masses in the detector frame.}
    \label{fig:m1_m2_posterior}
\end{figure}

The distribution of stationary and Gaussian noise in time domain can be characterized by a multivariate normal distribution, and the likelihood can be expressed as follows
\begin{equation}
\mathcal{L}(\boldsymbol{n}\,|\,\boldsymbol{\Sigma})=\frac{1}{\left(2\pi\right)^{k/2}\mathrm{det}\left(\boldsymbol{\Sigma}\right)^{1/2}}\exp\left(-\frac{1}{2}n_i \Sigma_{ij}^{-1}n_j\right).
\label{eq:td_likelihood_app}
\end{equation}
This likelihood is defined in the time domain, but it can also be calculated in the frequency domain by Fourier transforming the time-domain noise $\boldsymbol{n}$ into $\tilde{\boldsymbol{n}}$. We would expect both likelihood values to be consistent, provided that no approximations are introduced. However, in a typical GW data analysis, a window function is applied to the time domain data before it is Fourier transformed, and the likelihood will be computed using the approximated Whittle likelihood.

\begin{figure}[t]
    \centering
\includegraphics[width=1.0\linewidth]{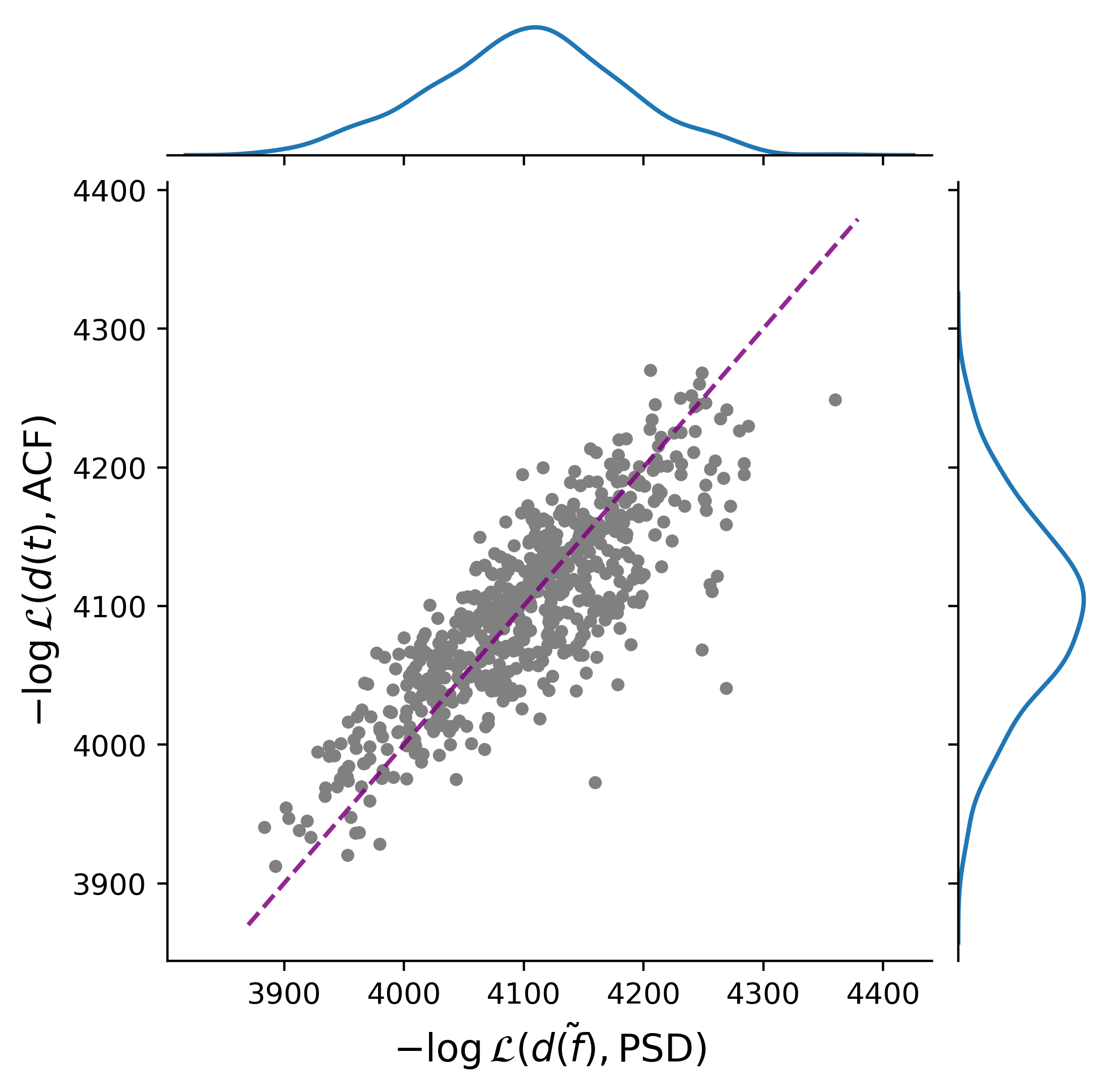}
    \caption{The log-likelihoods evaluated on 600 randomly generated 4s noise segments. For each random noise segment (gray scatter point), we compute the log-likelihood using two methods: one that directly utilizes Eq.~\ref{eq:td_likelihood} and another that employs the approximated Whittle likelihood in the frequency domain. The purple dashed line highlights the differences in likelihood calculations at these random noise segments. If the results from various methods are consistent, the distribution of the scatter points should closely align with this purple dashed line.}
    \label{fig:acf_psd_likelihood}
\end{figure}

In the context of single binary black hole parameter estimation for real GW data analysis, the SNR is typically quite high ($\rho \gtrsim 8$). \changed{In such cases, the use of the Whittle likelihood has become standard practice within the GW community, primarily due to its significantly faster evaluation during MCMC sampling. Although it is an approximation rather than an exact likelihood, any potential shortcomings or small biases introduced by this approximation are typically overlooked and generally considered negligible in practical analyses. We have tested this independently as follows.} In our mock data analysis, we inject a GW signal into a two-detector network of LIGO-Livingston, LIGO-Hanford with an joint SNR of $\rho = 17$ into the 4s long time-domain noise. We then perform the parameter estimation using different likelihoods with  the same  prior. Specifically, we begin by conducting coarse-level parameter estimation using the Whittle likelihood, followed by re-weighting all posterior samples using the full likelihood, as represented in Eq.~\ref{eq:td_likelihood}, which is computed in both the time domain, using the ACF, and frequency domain, using the $C_{ij}$) (refer to Section~\ref{subsec:finite_duration} and Eqs.~\ref{eq:td_likelihood}-\ref{eq:Cij} wherein). The posterior distribution for two of the intrinsic binary parameters is shown in Fig.~\ref{fig:m1_m2_posterior}, where $m_1$ and $m_2$ refer to the primary and secondary BBH masses in the detector frame. \changed{It is evident that there are no significant difference between the posteriors.}  

\changed{However, differences between likelihood computation methods cannot be ignored in analyses like ours, where the duty cycle is constructed by accumulating signal and noise evidence values from multiple segments — most of which either do not contain a signal or contain only weak signals. In Fig.\ref{fig:acf_psd_likelihood}, we show the case where no signals are injected.
We compare the log-likelihoods obtained from the analysis of 600 randomly selected $\unit[4]{s}$ noise segments generated in the time domain using two methods: (1) the likelihood, $\mathcal{L}\left(d(t),\mathrm{ACF}\right)$, computed directly using Eq.\ref{eq:td_likelihood} with the autocorrelation function (ACF); and (2) the likelihood, $\mathcal{L}\left(\tilde{d}(f),\mathrm{PSD}\right)$, calculated using the Whittle likelihood approximation with the true power spectral density (PSD). We see that there are subtle differences between the two, which becomes important in the context of our analysis.}

\changed{While Whittle likelihood is a good approximation for the analysis of loud signals, analysis like ours need precise measurement of the likelihood where the approximation has shortcomings. Furthermore, even though the time and frequency domain versions of the full likelihoods are equivalent, the time domain one is $>1000$ times more computationally efficient.}

\section{Further Insights into the Time-Reversed O3b Analysis}

\begin{figure}[t]
    \centering
\includegraphics[width=1.0\linewidth]{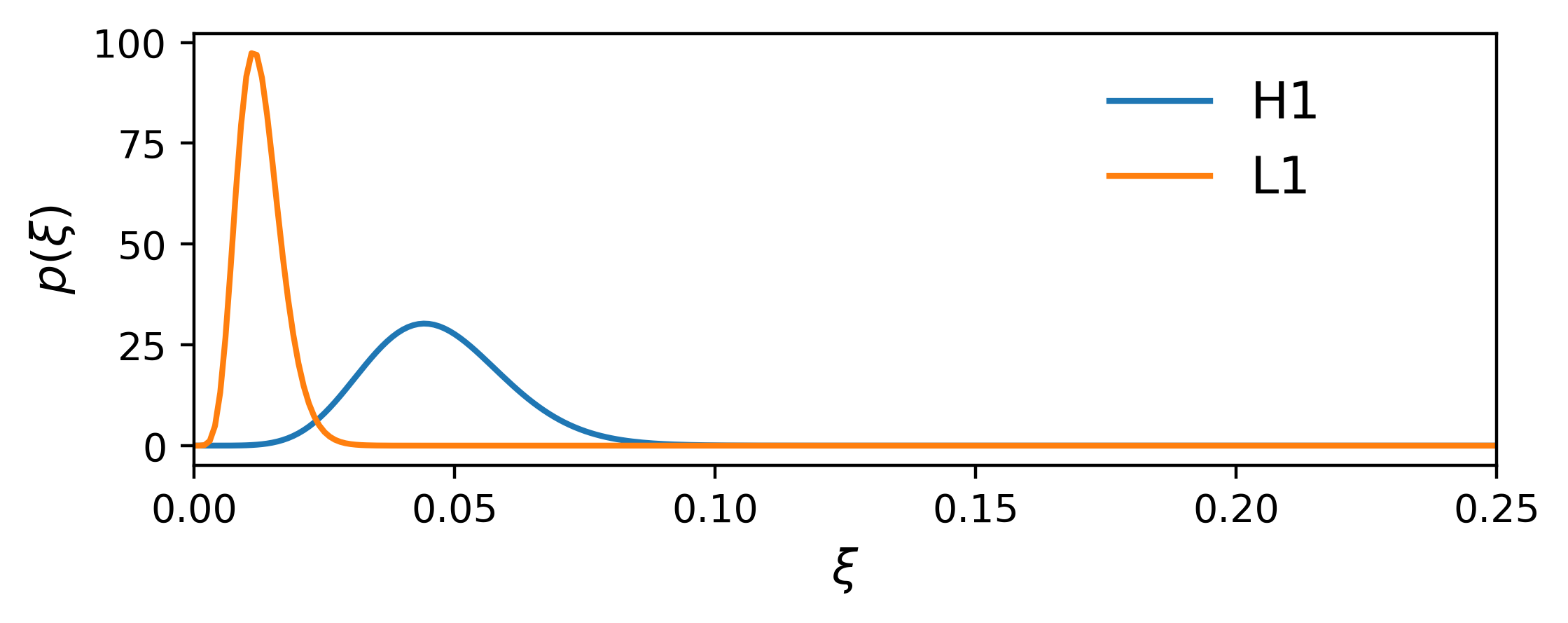}    \includegraphics[width=1.0\linewidth]{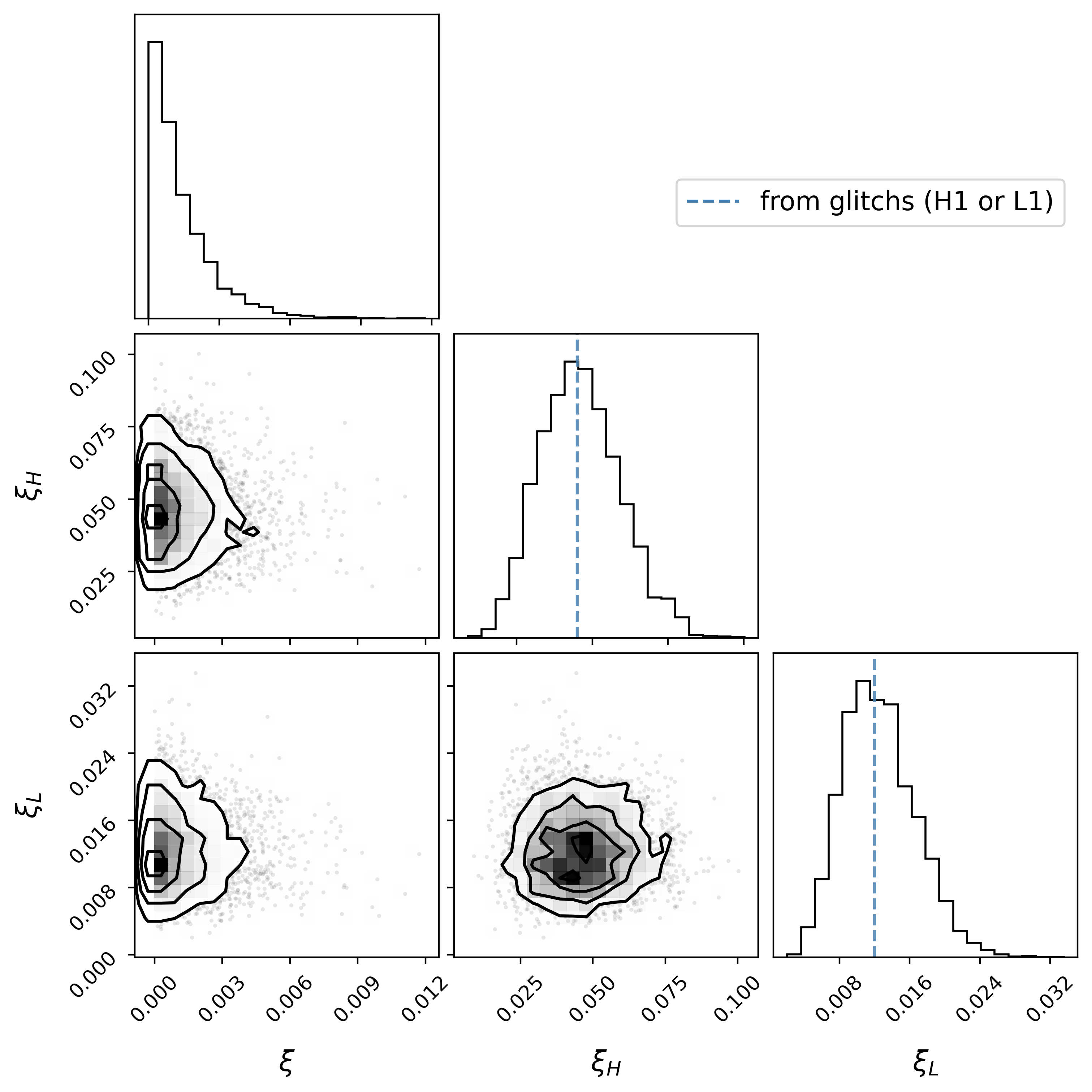}
    \caption{Top: Posterior distributions of the duty cycle parameter, $p(\xi\,|\,\mathcal{D})$, for individual detectors H1 and L1. They are obtained from the same data used in the top panel of Fig.~\ref{fig:real_H1L1}, but without employing the glitchy likelihood. Each segment is analyzed using the modified pipeline detailed in Figs.~\ref{fig:td_true_reweight} and~\ref{fig:draw_tbs_pipeline}. Bottom: Corner plot obtained from the parameter estimation of $\xi$, $\xi_H$ and $\xi_L$ utilizing the full mixture likelihood as described in Eq.~\ref{eq:glitch_likelihood}. This analysis incorporates glitches by collecting
    $\mathcal{Z}_S, \mathcal{Z}_N, \mathcal{Z}_g^{(1)}, \mathcal{Z}_g^{(2)}, \mathcal{Z}_N^{(1)}$ and $\mathcal{Z}_N^{(2)}$ for all segments as obtained based on the modified analysis framework shown in Fig.~\ref{fig:draw_tbs_pipeline}. The blue dashed lines represent the best-fit $\xi$ values, corresponding to their maximum likelihoods, obtained from the individual H1 and L1 detector analyses presented in the top panel.}
    \label{fig:real_glitches}
\end{figure}

In Section~\ref{sec:real-o3-noise}, we demonstrated that glitch contamination in the duty cycle $\xi$ can be mitigated by using the glitchy likelihood of Eq.~\ref{eq:glitch_likelihood}, where glitches are modeled as BBH-like waveforms occurring incoherently in the two detectors.  

Figure~\ref{fig:real_glitches} illustrates the importance of modeling glitches on the estimation of the astrophysical duty cycle. The top panel displays the posterior distributions of the duty cycle $\xi$ when individual detectors (H1 and L1) are analyzed without incorporating the full glitch mixture likelihood. These distributions reveal detector-specific non-zero peaks. In contrast, the bottom panel presents a corner plot of the astrophysical duty cycle $\xi$ and the glitch duty cycles $\xi_H$ and $\xi_L$ for the two detectors. While $\xi$ peaks at the true value of zero, both $\xi_H$ and $\xi_L$ exclude zero with more than 99\% credibility, Furthermore, the peak values for $\xi_H$ and $\xi_L$ obtained in this panel align closely with the respective peak values (blue dashed lines) of $\xi$ for H1 and L1 shown in the top panel. This alignment indicates that the individual detector biases observed earlier are effectively accounted for by the explicitly modeled weak glitch rates in each detector (we note that the data analyzed did not contain any loud glitches).  

Although the glitch duty cycles $\xi_H$ and $\xi_L$ may be considered nuisance parameters in the context of this study, their posteriors in Fig.~\ref{fig:real_glitches} highlight the importance of explicitly modeling incoherent glitches. Doing so is crucial for preventing glitch contamination in the astrophysical duty cycle $\xi$.

\section{Accounting for Selection Effects}\label{appendix:selection-effects}

\begin{figure}[t]
    \centering
    \includegraphics[width=1.0\linewidth]{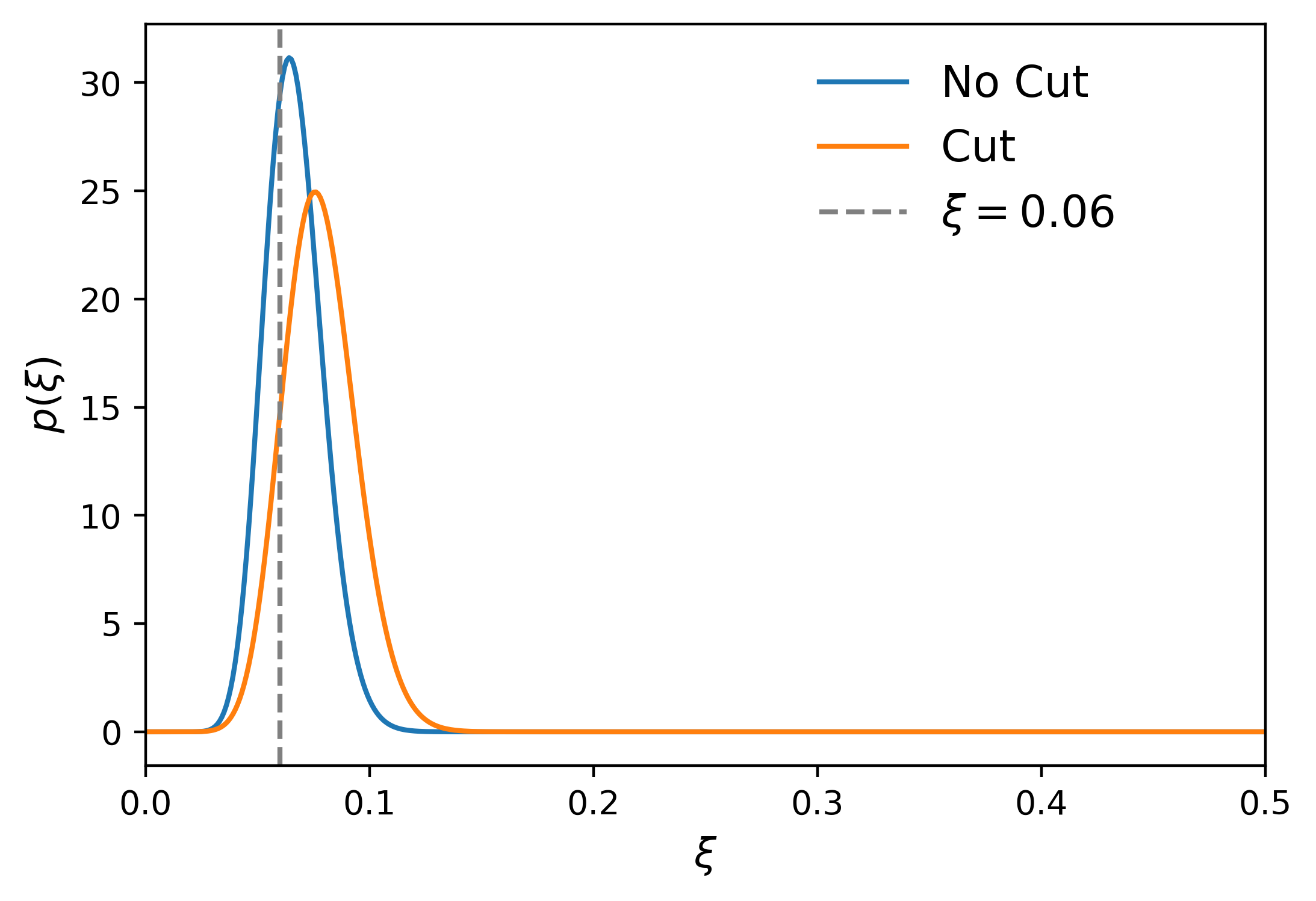}
    \caption{Posterior distributions for duty cycle $\xi$ with resolvable binaries (blue) and with resolvable binaries excluded (orange).
    Both posteriors are consistent with the true value $\xi=0.06$ represented by the dashed vertical line.
    However, excluding $\text{SNR}>8$ segments slightly broadens the posterior, which is calculated with Eqs.~\ref{eq:pdet} and ~\ref{eq:pdet-def} to take into account selection effects.
    The plot is made using 3800 4s segments.}
    \label{fig:tbs_select_func}
\end{figure}

In Eq.~\ref{eq:pdet} we introduce the factor $p_{\mathrm{det}}(\xi)$, which describes how the likelihood function is renormalized in the presence of selection effects. The variable $p_\text{det}$ represents the probability that a data segment is retained because it does \textit{not} have a single-detector, matched-filter signal-to-noise ratio $\text{SNR}<8$.
Let $f$ denote the probability that a binary black hole drawn from our signal model is detected with a matched filter $\text{SNR}>8$. This quantity can be estimated by generating a large number of signals from the prior and computing the fraction with $\text{SNR}>8$. 
The joint probability that any given segment contains a resolvable signal is therefore: $f \xi$.

Consequently, the probability that a segment does \emph{not} yield a detectable signal—that is, it survives the selection filter and is included in our sub-threshold analysis—is $1 - f\xi$. For $N$ such independent analysis segments, the probability that \textit{all} of them survive detection is given by\footnote{We have assumed here that no noise-only segments in the data will pass our selection cut.}

\begin{equation}
    p_{\mathrm{det}}(\xi) = \left(1 - f\xi \right)^N \,.
    \label{eq:pdet-def}
\end{equation}
 
In Fig.~\ref{fig:tbs_select_func}, we illustrate the impact of selection effects using a mock time domain dataset with an injected duty cycle of $\xi = 0.06$. We follow the flowchart in Fig.~\ref{fig:draw_tbs_pipeline} to analyze each segment, addressing both PSD uncertainty and finite-duration effects. The blue curve corresponds to the case where no SNR threshold is applied, so no selection correction is needed. The orange curve shows the posterior after removing $\text{SNR}>8$ segments and correcting for selection effects using Eqs.~\ref{eq:pdet} and ~\ref{eq:pdet-def} as part of a post-processing step. Both posteriors are consistent with the true value of $\xi$ indicated by the dashed vertical line.
By excluding events with $\text{SNR}>8$, the orange posterior is slightly broader.


\bibliography{main}

\end{document}